\documentclass[12pt]{article}
\pdfoutput=1
\usepackage{amsmath}
\usepackage{amsfonts}
\usepackage{amssymb}
\usepackage{graphicx}
\providecommand{\U}[1]{\protect\rule{.1in}{.1in}}

\begin{document}

\title{A Microscopic Cluster Description of $^{12}C$.}

\author{V. S. Vasilevsky$^1$, \ F. Arickx$^2$, W. Vanroose$^2$, J. Broeckhove$^2$\\$^1$Bogolyubov Institute for Theoretical Physics, Kiev, Ukraine\\$^2$Departement Wiskunde-Informatica, Universiteit Antwerpen, \\Antwerpen, Belgium}

\maketitle

\date{}

\begin{abstract}
We investigate both bound and resonance states in $^{12}C$ embedded in a three-$\alpha$-cluster
continuum using a three-cluster microscopic
model. The model relies on the Hyperspherical Harmonics basis to enumerate the
channels describing the three-cluster discrete and continuous spectrum
states. It yields the most probable distribution of the three $\alpha$-clusters in space, and the dominant
decay modes of the three-cluster resonances. 
\end{abstract}

\section{Introduction}

The $^{12}C$ nucleus is an interesting example of the so-called Borromean nuclei, as it has no bound states
in any two-cluster subsystem of its three-cluster configuration.
The lowest dissociation threshold (7.276 MeV above the ground state) is that of a three $\alpha$ particles
disintegration. This three-cluster configuration is thus responsible to a great
extent for the formation of a few bound, and many resonance
states. The next threshold is of a two-cluster nature: $^{11}B+p$ \cite{1990NuPhA.506....1A}.
It opens when the excitation energy of $^{12}C$
exceeds 15.96 MeV. One therefore expects only a negligible influence
of the latter channel on the bound and resonance states of $^{12}C$ in the vicinity of the
$\alpha+\alpha+\alpha$ threshold.

The $^{12}C$ nucleus is unique because of its excited ``Hoyle state''. This state is important in the context
of the nucleosynthesis of carbon in helium-burning red giant stars. It
is a $0^{+}$ state with an energy of 7.65 MeV above the ground state, or 0.4
MeV above the three-cluster $\alpha+\alpha+\alpha$ threshold. Its width is
only 8.5 eV, indicating a long lifetime. One immediately
relates this to the $0^{+}$ state in $^{8}Be$ described by two
$\alpha$ particles, with an energy of 0.092 MeV above the $\alpha+\alpha$ threshold,
and a width of 5.57 eV.

Many efforts have been made to reproduce the experimentally observed structure of $^{12}C$, and
to explore and understand the nature of the ground, excited and resonance states. This was e.g.\ done
within so-called semi-microscopic models (considering structureless $\alpha$-particles)
\cite{2004PhRvC..70a4006F,2005JPhG...31.1207F,
2007EPJA...31..303A, 2008JPhCS.111a2017A,
2008PhRvC..77f4305A, 2010PhRvL.105b2501C,
2011PhLB..695..324D} and within fully microscopic models 
\cite{1997NuPhA.618...55P,2004NuPhA.738..455K,2004NuPhA.738..495F,2004PhRvC..69c7002F,
2004FBS....34..237F,2002PThPh.107..745F,2004PhRvC..70b4002F,2005PhRvC..71b1301K,
2006PhRvC..74f4311A,2007PhRvC..76e4003T,2007NuPhA.792...87K,1987PhRvC..36...54D,
2008JPhCS.111a2045O,2008PhLB..659..160S,
2010JPhG...37f4010D,2011PhRvC..83b4301Y}.

A rather general feature of the calculations is that, with potentials which
adequately reproduce the $\alpha$-$\alpha$ interaction (this includes the phase
shifts for $0^{+}$, $2^{+}$ and $4^{+}$ states, and the position of the
corresponding resonance states), one obtains a noticeably overbound ground state for $^{12}C$.

To determine the energies and widths of the resonance
states created by a three-cluster continuum, only a few methods can be used. One popular method
for obtaining the resonance properties in many-cluster, many channel systems
is the Complex Scaling Method (see reviews \cite{1998PhR...302..212M,
1983PhR....99....1H} and references therein).
Other methods start from a calculated form of the $S$-matrix in a wide energy range, and
determine the resonance states as the pole(s) of the $S$-matrix. The advantage of these methods is that they
provide the scattering quantities (such as phase-shifts, cross-sections, \ldots) and
the resonance properties (energies and widths), as well as the wave functions of scattering and resonance states.
The latter then allow one to obtain more information about the nature of the resonance
states. 

$^{12}C$ is known from theory and experiment (see, e.g., \cite{2010JPhG...37j5104M} and \cite{2009PhRvC..80d1303F})
to have some very narrow resonances above the three $\alpha$ threshold.
One may wonder why a system with several open channels does not decay instantly, but 
manifests these narrow resonance states. There are two possible
answers to this question. First, a resonance state appears  in one single
channel of the multi-channel system. Such particular channel is usually weakly
coupled to a number, or all, of the other open channels. It is well-known that this weak coupling of channels
predetermines the existence of long-lived resonance states. Second, a resonance
can be more or less uniformly distributed over all open channels, and the compound
system needs (some) time for the resonance to be accumulated by one or a few number of
open channels in order to decay into. Such a distribution over many open channels
leads to very narrow resonances,  as was predicted
by A. Baz' \cite{1976JETP...43..205B}. It is referred to as diffusion-like
processes in scattering. This type of resonance is attributed to the effect that  ``the system spends
most of its time wandering from one channel to another'' \cite{1976JETP...43..205B}. 

In this paper we wish to calculate and analyze the bound and continuum structure of $^{12}C$,
and gain some insight in the nature of these states. 
Indeed, in some publications (e.g.\
\cite{1995PhRvC..51..152A,1992NuPhA.549..431M,2007PThPh.117..655K,2004NuPhA.738..357N})
the suggestion for a dominant linear,
chain-like, three-cluster structure appears for some of the $^{12}C$ resonances.
We will look for confirmation of this structure.
To this end, we determine the
most probable configuration of the three $\alpha$ particles both in coordinate and
momentum space. We also qualify those channels on which the resonance
states of $^{12}C$ preferentially decay. 

The main results of this paper are obtained by applying the ``Algebraic Model in a Hyperspherical Harmonics Basis'' (AMHHB)
\cite{2001PhRvC..63c4606V,2001PhRvC..63c4607V, 2007JPhG...34.1955B} on a configuration
of three $\alpha$-particles. In this model the three clusters are treated equally, and their
relative motion descrybed by Hyperspherical Harmonics. The latter enumerate the channels of the three-cluster continuum
and allow to implement the correct boundary conditions for the three-cluster exit channels.
The AMHHB has been applied successfully to study resonances in nuclei with a large excess of protons or neutrons
such as $^{6}He$, $^{6}Be$, $^{5}H$. The method provides
the energies and widths of the resonances, and their total and partial widths, as well as
the corresponding wave functions. The latter allow to analyze the nature
of the resonance states.
The results of this model are compared to those obtained in other, more or less comparable, microscopic descriptions
from the literature, and to experiment.

In the next section we elaborate on the method used to calculate
the spectrum of $^{12}C$.
Section three focuses on the results obtained
using this method. We
also present correlation functions and density functions to characterize
more precisely the spatial configuration of the three $\alpha$ particles for specific resonance states.
We also compare the results to those of
other microscopic calculations as well as to experiment.

\section{The microscopic cluster model}

In this section we describe the microscopic
model used to determine the structure of $^{12}C$ in the present paper.
As it has already been introduced and used in several
publications, we will limit ourselves to the most important notations and aspects
of importance to the current calculations.

\subsection{The three-cluster AMHHB model}

The three-cluster ``Algebraic Model in a Hyperspherical Harmonics Basis'' (AMHHB) \cite{2001PhRvC..63c4606V, 2001PhRvC..63c4607V,A4Resonances2004, 2007JPhG...34.1955B} will be applied to a single $^{12}C=\alpha+\alpha+\alpha$
three-cluster configuration.

This model takes a Hyperspherical Harmonics basis (HH) to
characterize and enumerate the different three-cluster channels. In each of these channels an
oscillator basis describes the radial behavior, and is used to expand the many-particle wave function.
A matrix version of the Schr\"odinger equation is obtained after substitution of this wave function.
It solved by the Algebraic Method (also called the Modified J-Matrix method
\cite{2007JPhG...34.1955B}) for both bound and scattering states using the correct asymptotics.

A similar approach, using the Hyperspherical Harmonics, was proposed in
\cite{2004NuPhA.740..249K,2009PhRvC..80d4310D} in coordinate representation,
using the generator coordinate technique to solve the corresponding Schr\"odinger equation.

The AMHHB wave function for $^{12}C$ is written as
\begin{eqnarray}
\Psi & = &\widehat{\mathcal{A}}\left\{  \Phi\left(  \alpha_{1}\right)\Phi\left(  \alpha_{2}\right)  \Phi\left(  \alpha_{3}\right)  f\left(\mathbf{x},\mathbf{y}\right)  \right\} \label{eq:001}\\
& = &\widehat{\mathcal{A}}\left\{  \Phi\left(  \alpha_{1}\right)  \Phi\left(\alpha_{2}\right)  \Phi\left(  \alpha_{3}\right)  f\left(  \rho,\theta;\widehat{\mathbf{x}},\widehat{\mathbf{y}}\right)  \right\} \nonumber \\ 
& = &\sum_{n_{\rho},K,l_{1},l_{2}}C_{n_{\rho},K,l_{1},l_{2}}\left\vert n_{\rho},K,l_{1},l_{2};LM;\left(  \rho,\theta;\widehat{\mathbf{x}},\widehat{\mathbf{y}}\right)  \right\rangle \nonumber
\end{eqnarray}
where $\left\vert n_{\rho},K,l_{1},l_{2};LM\right\rangle $ is a cluster
oscillator function \cite{2001PhRvC..63c4606V}:
\begin{eqnarray}
&&\left\vert n_{\rho},K,l_{1},l_{2};LM\right\rangle = \\ \label{eq:002}
&&\ \ \ \ \ \widehat{\mathcal{A}}\left\{  \Phi\left(  \alpha_{1}\right)  \Phi\left(
\alpha_{2}\right)  \Phi\left(  \alpha_{3}\right)  R_{n_{\rho},K}\left(
\rho\right)  \chi_{K,l_{1},l_{2}}\left(  \theta\right)  \left\{  Y_{l_{1}%
}\left(  \widehat{\mathbf{x}}\right)  Y_{l_{2}}\left(  \widehat{\mathbf{y}%
}\right)  \right\}  _{LM}\right\} \nonumber
\end{eqnarray}
These functions are enumerated by the number of hyperradial excitations $n_{\rho}$,
hyperspherical momentum $K$ and two partial orbital momenta $l_{1},l_{2}$.
The vectors $\mathbf{x}$ and $\mathbf{y}$ form a set of Jacobi coordinates, and $\rho$\ and
$\theta$ are hyperspherical coordinates, related to the Jacobi vectors by:
\begin{eqnarray}
\rho =\sqrt{\mathbf{x}^{2}+\mathbf{y}^{2}}\nonumber\\
\left\vert \mathbf{x}\right\vert  =\rho\cos\theta,\quad\left\vert
\mathbf{y}\right\vert =\rho\sin\theta 
\end{eqnarray}
The notation $\widehat{\mathbf{x}}$ and $\widehat{\mathbf{y}}$ refers to unit length vectors.
Vector $\mathbf{x}$ corresponds to the distance between two selected
$\alpha$ particles, with an associated partial orbital angular momentum $l_{2}$. Vector $\mathbf{y}$ is the displacement of the third
$\alpha$ particle with respect to the center of mass of the other two, with an associated angular momentum $l_{1}$.
The three quantum
numbers $c=\left\{  K,l_{1},l_{2}\right\}$ determine the channels of the three-cluster system in the AMHHB.

The fact that all three clusters are identical leads to some specific issues. The
wave function (\ref{eq:001}) for $^{12}C$ is antisymmetric with respect to the permutation of any
pair of nucleons. Because the three clusters are identical, this function should be symmetric
with respect to the permutation
of any pair of alpha particles. This imposes constraints on the allowed quantum
numbers of the wave function. Due to this symmetry, for instance, the
partial orbital momentum $l_{2}$ of a two-cluster subsystem can only have even
values. As the parity of $^{12}C$ states is defined as $\pi=\left(  -1\right)  ^{l_{1}+l_{2}}$,
it is fully determined by the partial orbital angular momentum $l_{1}$ of
the relative motion of the remaining cluster with respect to the two-cluster subsystem.

In \cite{2002PThPh.107..993F} and \cite{2004PhRvC..69c7002F} it was suggested to
use a symmetrization operator to construct the proper basis states. For a discussion on the symmetry of a system
with three identical clusters we refer to
\cite{2009NuPhA.826...24L}.

The symmetrical Hyperspherical Harmonics basis for a
three-particle system was realized many years ago
(see, e.g., \cite{1965AnPhy..35...18Z, Nyiri:1979qm}). An explicit form of a
few basis functions for small values of the total angular
momentum ($L=0, 1$ and 2) can be derived. However it is extremely intricate
to use for explicit calculation of matrix elements.

An alternative approach to obtain
such matrix elements without an explicit realization of the basis functions consists in using the generating function technique.
One can indeed construct a generating function for the overlap and hamiltonian kernels of
$^{12}C$, using the procedure explained in
\cite{2001PhRvC..63c4606V}, that satisfies all required symmetry conditions, including the cluster symmetric permutation behavior.
Explicit matrix elements of the
operators
can then be obtained by using recurrence relations.
The standard approach in the AMHHB is to extract matrix elements characterized by explicit $l_1,l_2$ quantum numbers.
These, however,
do not yet correspond to the desired symmetrical Harmonics. Indeed, the states $\left\vert n_{\rho},K,l_{1},l_{2};LM\right\rangle $
for fixed $n_\rho$ and $K$ do not belong to the desired symmetrical
irreducible representation of S(3), the permutation group of the three $\alpha$ clusters, with Young tableau [3].
They are, in fact, linear combinations of the Young tableau [3] and the non-symmetrical Young tableaus [2,1] and [111].

The antisymmetrization operator in the standard AMHHB basis has non-zero matrix elements 
\begin{equation}
\left\langle n_{\rho},K,l_{1},l_{2};LM\left\vert \widehat{A}\right\vert
\widetilde{n}_{\rho},\widetilde{K},\widetilde{l}_{1},\widetilde{l}%
_{2};LM\right\rangle,
\end{equation}
for fixed oscillator shells with $N_{sh}$=$2\ n_{\rho
}+K=2\ \widetilde{n}_{\rho}+\widetilde{K}$. By selecting only the matrix elements with
hyperradial quantumnumber $n_{\rho}=\widetilde{n}_{\rho}$ and
hypermomentum $K=\widetilde{K}$, one obtains relatively small matrices
whose eigenfunctions $\left\vert n_{\rho},K,\nu;LM\right\rangle$ with non-zero eigenvalues are of the correct
symmetrical Hyperspherical type, due to the symmetry properties of the generating function.
This procedure 
is similar to the procedure of obtaining the Pauli allowed states
in three-cluster systems (for details see \cite{2010PPN....41..716N}).

\begin{figure}[!ht]
	\begin{center}
		\includegraphics[width=0.85\linewidth]{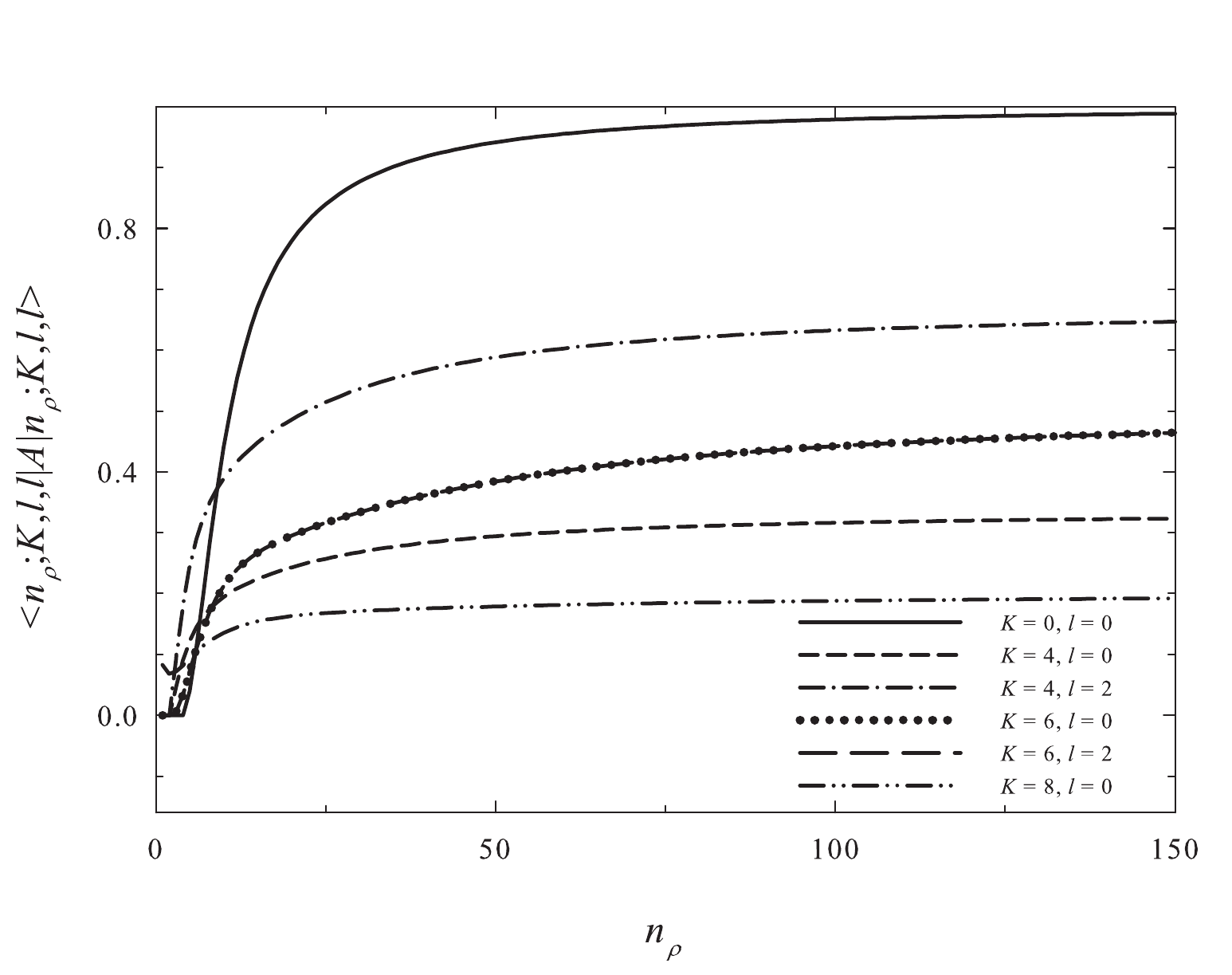}
		\caption{Matrix elements of the antisymmetrization operator in the unsymmetrized Hyperspherical basis.}
		\label{Fig:Overlap_orig_L0}
	\end{center}
\end{figure}

This is demonstrated in Figure \ref{Fig:Overlap_orig_L0} where the diagonal matrix elements of the antisymmetrization operator between the original
Hyperspherical Harmonics are displayed for total angular momentum $L=0$, for all channels up to $K=8$.
One notices that the matrix elements
\begin{equation}
\left\langle n_{\rho},K,l_{1}%
=l_{2};L=0\left\vert \widehat{A}\right\vert n_{\rho},K,l_{1}=l_{2}%
;L=0\right\rangle \label{eq:006}
\end{equation}
do not tend to unity, as one could expect, but to
some fixed values. Analysis shows that these asymptotic values of (\ref{eq:006}) correspond to the weights of the symmetrized
Hyperspherical Harmonics with Young tableau [3], within the original Harmonic. 

\begin{figure}[!ht]
	\begin{center}
		\includegraphics[width=0.85\linewidth]{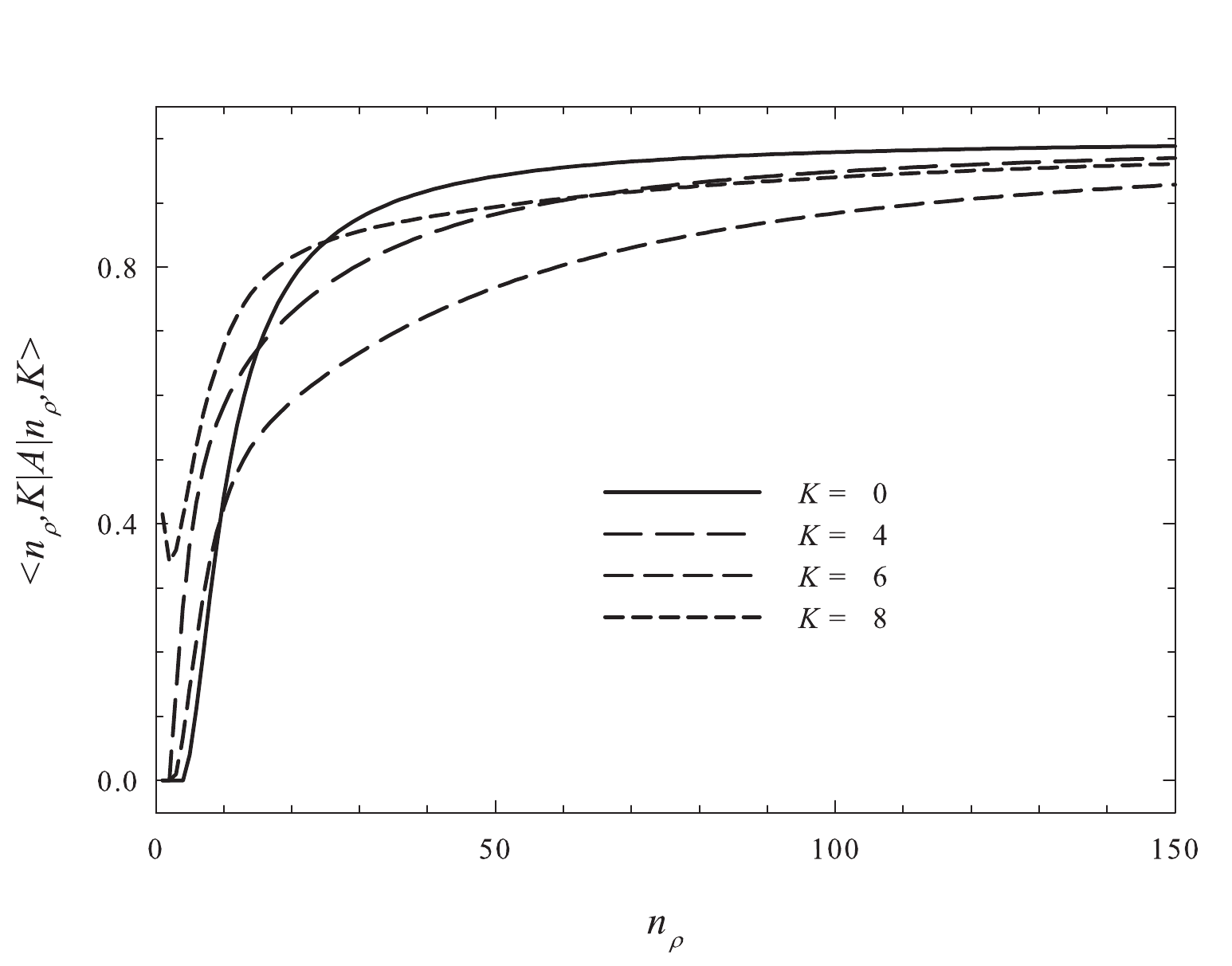}
		\caption{Matrix elements of the antisymmetrizator in the symmetrized Hyperspherical basis.}%
		\label{Fig:Overlap_symm_L0}%
	\end{center}
\end{figure}

The eigenvalues obtained after diagonalization however, which are matrix elemts of symmetrized Harmonics, do display
the correct asymptotic behavior, i.e.\ all tend to unity, as can be seen in Figure
\ref{Fig:Overlap_symm_L0}.

In Table \ref{Tab:SymmNonsymmChs} we display both the total number of
(original) nonsymmetrized and of symmetrized channels for different values of
the total orbital momentum. The symmetrization significantly
reduces the number of channels compatible with the maximal value of
Hypermomentum $K_{\max}$. Only even
values of the partial orbital momentum $l_{2}$ are considered because of the symmetry rules for two-cluster subsystems.

\begin{table}[!ht] \centering
\caption{Number of channels for unsymmetrized and symmetrized Hyperspherical
Harmonics.}%
\begin{tabular}
[c]{|c||c|c|c|c|c|}\hline
$J^{\pi}$ & $0^{+}$ & $2^{+}$ & $4^{+}$ & $1^{-}$ & $3^{-}$\\\hline\hline
$K_{\max}$ & 14 & 14 & 14 & 13 & 13\\\hline
$N_{ch}$ ($\left\{  K,l_{1},l_{2}\right\}  $) & 20 & 44 & 54 & 28 & 42\\\hline
$N_{ch}$ ($\left\{  K,\nu\right\}  $) & 8 & 16 & 19 & 9 & 14\\\hline
\end{tabular}
\label{Tab:SymmNonsymmChs}
\end{table}

In \cite{2001PhRvC..63c4606V} and \cite{2001PhRvC..63c4607V} the importance  and meaning of the effective charges
$Z_{c,\widetilde{c}}$ were defined in the context of the AMHHB, and explicitly discussed for the $^{6}Be$ nucleus in
three-cluster configuration $^{4}He+p+p$. The effective charge
determines the asymptotic form of the three-cluster potential originating from the
Coulomb interaction, which has the form
\begin{equation}
V_{c,\widetilde{c}}^{(C)}=\frac{Z_{c,\widetilde{c}}}{\rho}%
\label{eq:Coul_asymp}%
\end{equation}
It was shown that it is of crucial importance for implementing the correct boundary conditions for
the three-cluster continuum states. 

The symmetrization influences the behavior of the effective charges. In Table \ref{Tabl:ZeffOrig} we display the effective charges for the $0^{+}$ state of
$^{12}C$, calculated in the original, nonsymmetrized, basis of the
Hyperspherical Harmonics, for $K_{max} = 8$. One easily verifies that they coincide
with those calculated in \cite{2010JPhG...37f4010D}.%
\begin{table}[!ht] \centering
\caption{Effective charges for the $J^{\pi} = 0^{+}$ state of $^{12}C$.}%
\begin{tabular}[c]{|r||r|r|r|r|r|r|r|r|}\hline
$\left(  K,l_{1},l_{2}\right)  $ & (0,0,0) & (4,0,0) & (4,2,2) & (6,0,0) &
(6,2,2) & (8,0,0) & (8,2,2) & (8,4,4)\\\hline \hline
(0,0,0) & 28.81 & 2.47 & 3.49 & 2.74 & -2.74 & 0.87 & 0.00 & 1.04\\ \hline
(4,0,0) & 2.47 & 32.157 & -1.13 & 3.95 & -0.31 & 4.67 & 0.00 &
1.95\\\hline
(4,2,2) & 3.49 & -1.13 & 31.35 & 1.72 & -4.30 & 0.00 & 0.66 & 0.00\\\hline
(6,0,0) & 2.74 & 3.95 & 1.72 & 33.48 & -2.51 & 4.63 & 0.00 &
0.45\\\hline
(6,2,2) & -2.74 & -0.31 & -4.30 & -2.51 & 34.29 & 0.00 & 0.62 &
0.00\\\hline
(8,0,0) & 0.87 & 4.67 & 0.00 & 4.63 & 0.00 & 34.29 & 0.00 & -2.38\\\hline
(8,2,2) & 0.00 & 0.00 & 0.66 & 0.00 & 0.62 & 0.00 & 33.08 & 0.00\\\hline
(8,4,4) & 1.04 & 1.95 & 0.00 & 0.45 & 0.00 & -2.38 & 0.00 & 32.41\\\hline
\end{tabular}
\label{Tabl:ZeffOrig}
\end{table}

In Table \ref{Tabl:ZeffSymm} we display the effective charges in the symmetrized basis.
Only four channels remain after symmetrization. In particular no $K=2$ channel remains,
so we ommitted these also in Table \ref{Tabl:ZeffOrig}
even though they have a non-zero contribution.
\begin{table}[!ht] \centering
\caption{Effective charges for the $J^{\pi} = 0^{+}$ state of $^{12}C$ for
symmetrized channels}%
\begin{tabular}
[c]{|r||r|r|r|r|}\hline
$\left(  K,v\right)  $ & (0,1) & (4,1) & (6,1) & (8,1)\\\hline\hline
(0,1) & 28.810 & 4.277 & 3.880 & 1.139\\\hline
(4,1) & 4.277 & 30.556 & 5.217 & 2.301\\\hline
(6,1) & 3.880 & 5.217 & 35.990 & 1.457\\\hline
(8,1) & 1.139 & 2.301 & 1.457 & 31.450\\\hline
\end{tabular}
\label{Tabl:ZeffSymm}
\end{table}

It goes without saying that the asymptotic form of the
effective three-cluster potential which originates from the nucleon-nucleon
interaction \cite{2001PhRvC..63c4606V}%
\begin{equation}
V_{c,\widetilde{c}}^{(NN)}=\frac{V_{c,\widetilde{c}}}{\rho^{3}}%
\label{eq:NN_asymp}%
\end{equation}
is also influenced by the symmetrization.
This asymptotic component is very important for obtaining the correct values of
the $S$ matrix. We do not dwell on its explicit form here, but apply a procedure similar to that for the effective charges.

\subsection{Phases, Eigenphases and Resonances}

After solving the system of linear equation of the AMHHB model,
we obtain the wave functions of the continuous
spectrum states, and the scattering $S$-matrix. We consider two
different representations of the $S$-matrix.

In the first representation, the elements of the $S$-matrix
are described through the phase shifts $\delta_{ij}$ and the inelastic
parameters $\eta_{ij}$:%
\begin{equation}
S_{ij}=\eta_{ij}\exp\left(  2i\delta_{ij}\right) \label{eq:031}%
\end{equation}
of which one usually only analyzes the diagonal matrix elements by displaying
the $\delta_{ii}$ and $\eta_{ii}$ quantities. In the second representation the $S$-matrix is
reduced to diagonal
form, leading to the so-called eigenphases, which now represent the elastic scattering of the
many-channel system in independent (uncoupled) eigenchannels:
\begin{equation}
\left\Vert S\right\Vert =\left\Vert U\right\Vert ^{-1}\left\Vert D\right\Vert
\left\Vert U\right\Vert \label{eq:032}%
\end{equation}
Here $\left\Vert U\right\Vert $ is an orthogonal matrix, connecting
both representations, and
$\left\Vert D\right\Vert $ is a diagonal matrix with nonzero elements%
\begin{equation}
D_{\alpha\alpha}=\exp\left(  2i\delta_{\alpha}\right) \label{eq:033}%
\end{equation}
defining the eigenphases $\delta_{\alpha}$.

The phases shifts $\delta_{ii}$, inelastic parameters $\eta_{ii}$ and
eigenphases $\delta_{\alpha}$ then provide sufficiently detailed information about
the channels that are
involved in the production of resonance states. The eigenphases are used to
extract the resonance positions and total widths in
the traditional way
\begin{equation}
  \left. \frac{d^{2}{\delta_{\alpha}} }{dE^{2}}\right\vert _{E=E_{r}}=0,
  \quad\Gamma=2\left( \left. \frac{d\delta_{\alpha}}{dE}\right\vert _{E_{r}}\right)  ^{-1}
  \label{eq:phshanalysis}
\end{equation}
whereas
the orthogonal matrix $\left\Vert U\right\Vert $ leads to the
partial decay widths of the resonance (for details see, e.g., \cite{2007JPhG...34.1955B}).

\subsection{Correlation functions and density distributions.}

As we pointed out, the AMHHB model allows to calculate the scattering properties, but also to
obtain the obtain wave function at any energy, in particular at the resonance positions.
The latter is of the utmost importance to analyze the nature of the system at these energies.

Within the AMHHB model the solution is fully expressed by the expansion coefficients
$\left\{  C_{n_{\rho},c}\right\} $ and the $S$-matrix.
The expansion coefficients $\left\{  C_{n_{\rho},c}\right\} $ determine
both the total three-cluster wave function of a compound system $\Psi$, as well as the wave function of the relative motion of three clusters $f\left(  \mathbf{x},\mathbf{y}\right) $ (see eq.\ (\ref{eq:001})).

The latter contains all information on the dynamic behavior of the three-cluster system for
bound as well as continuum states. It is interesting to note that these coefficients are identical
in both the representations of the wave function in coordinate and momentum space, because
of the Fourier transform properties of the oscillator states. The wave function
$f\left( \mathbf{k},\mathbf{q}\right) $ in momentum space has arguments that are directly
related to the coordinate representation: $\mathbf{k}$ is the momentum of relative motion
of two clusters, whereas $\mathbf{q}$ is the momentum of the third cluster with respect to the
center of mass of the two-cluster subsystem.

We obtain the density distribution in coordinate space as
\begin{equation}
D\left(  x,y\right)  =D\left(  \rho,\theta\right)  =\int\left\vert f\left(
\mathbf{x},\mathbf{y}\right)  \right\vert ^{2}d\widehat{\mathbf{x}}%
~d\widehat{\mathbf{y}}\label{eq:041}%
\end{equation}
and the corresponding correlation function as%
\begin{equation}
C\left(  x,y\right)  =C\left(  \rho,\theta\right)  =x^{2}y^{2}\int\left\vert
f\left(  \mathbf{x},\mathbf{y}\right)  \right\vert ^{2}d\widehat{\mathbf{x}%
}~d\widehat{\mathbf{y}}\label{eq:042}%
\end{equation}
directly from the wave function of relative motion $f\left(  \mathbf{x},\mathbf{y}\right) $.
Both the density distribution and correlation function in momentum space are obtained in the same way
using the wave function of relative motion in momentum space
$f\left( \mathbf{k},\mathbf{q}\right) $.

In a calculation with $N_{ch}$ open channels, one obtains $N_{ch}$ independent
wave functions describing the elastic and inelastic processes in the many-channel
system. It is quite impossible to analyze all of these wave functions when many channels
are open. Some principles have to be set up on how to select the most important wave functions.
In \cite{2007JPhG...34.1955B} we formulated some criteria for selecting the dominant wave
function of a resonance. We will use the same criteria in this paper to select the ``resonance
wave functions''.

\section{Calculations and results}

In the present calculations for $^{12}C$ we consider for the nucleon-nucleon interaction
the Minnesota potential \cite{kn:Minn_pot1}.
The oscillator basis is characterized by an
oscillator length $b=1.2846$ fm, to minimize the ground state
energy of the $\alpha$ particle using the above potential.

Parameter $u$ of the Minnesota potential is
taken to be $u=0.94$ in order to reproduce the phase shifts for $\alpha
+\alpha$ scattering, and the $0^{+}$, $2^{+}$ and $4^{+}$ resonances in $^{8}Be$.
The same parameters were used by Arai \cite{2006PhRvC..74f4311A}.

The $^8Be=\alpha+\alpha$ two-cluster substructure is of key importance in the description of $^{12}C$.
We present $\alpha+\alpha$ resonance properties in Table \ref{Tabl:8BeResonances}.
\begin{table}[!ht] \centering
\caption{Resonance properties for $^8Be$ obtained with different methods.}
\begin{tabular}
[c]{|c|r|r|r|r|}\hline
& \multicolumn{2}{|c}{AMOB} & \multicolumn{2}{|c|}{Arai \cite{2006PhRvC..74f4311A}}\\\hline
$J^{\pi}$ & $E$, MeV & $\Gamma$, keV & $E$, MeV & $\Gamma$, MeV\\\hline
$0^{+}$ & 0.022 & 6.30 10$^{-10}$ & 0.03 & $<$10$^{-6}$\\
$2^{+}$ & 2.93 & 1.51 & 2.9 & 1.4\\
$4^{+}$ & 12.55 & 5.01 & 12.5 & 4.8\\\hline
\end{tabular}
\label{Tabl:8BeResonances}
\end{table}
The AMOB model takes a set of oscillator functions to describe the intercluster behavior, and the Algebraic Model 
to obtain the phase shifts for $\alpha+\alpha$ scattering (see e.g.\ \cite{2005PhRvC..71d4322S}).
We include a comparison to the results of 
of Arai in his paper on $^{12}C$ \cite{2006PhRvC..74f4311A}, where he uses the ``analytical continuation of the $S$ matrix to the complex plane'' method with Complex Scaling to obtain the resonance characteristics.

These results form a first test of the consistency of the different expansion methods used, applied to the two-cluster subsystem.
Although quite similar, one still notices that the resonance properties of the two-cluster
$\alpha-\alpha$ system have a slight dependence on the method used.

\subsection{The Potential and Coulomb interaction in AMHHB}

\begin{figure}[!htb]
\begin{center}
\includegraphics[width=0.85\linewidth]{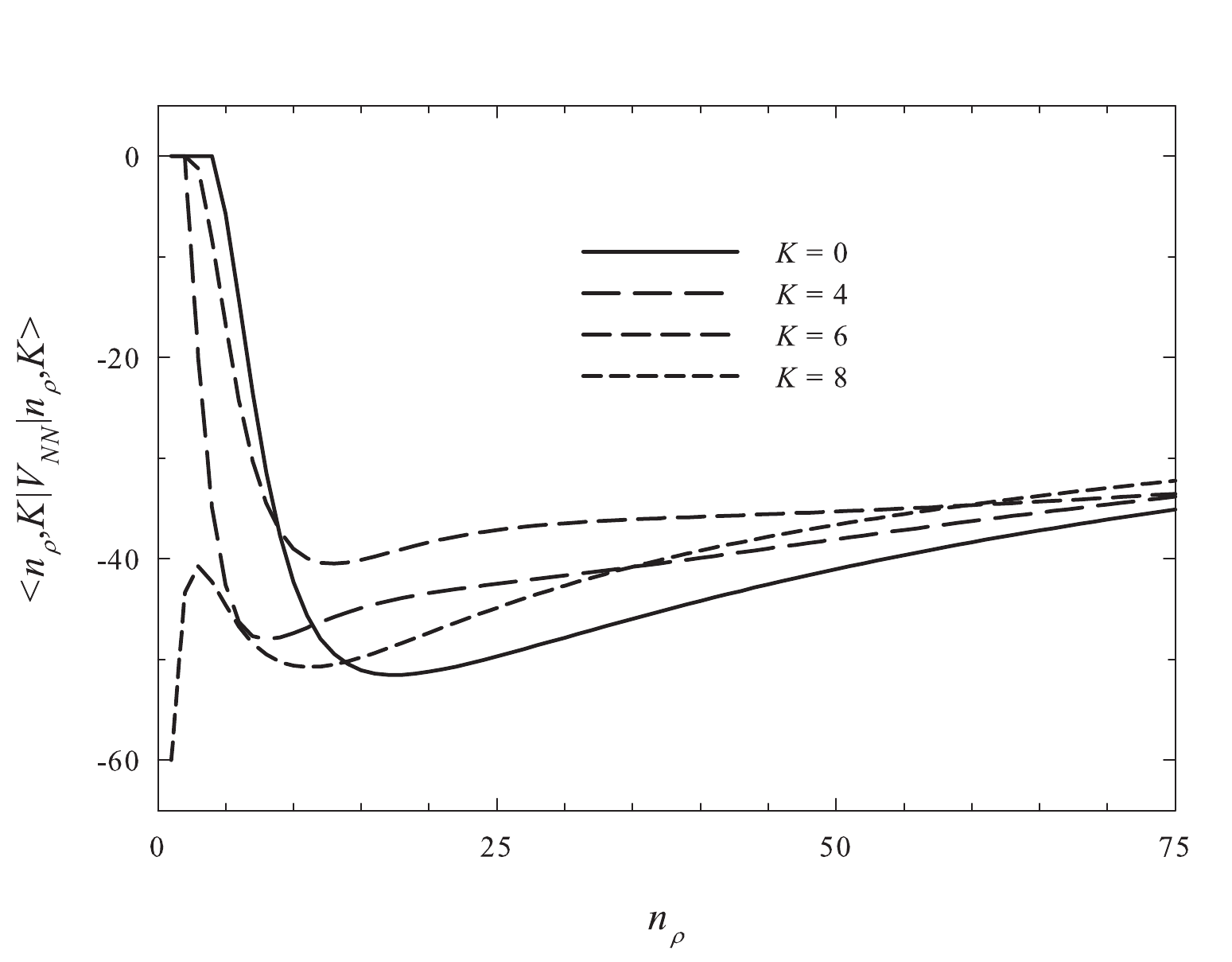}
\caption{Diagonal matrix elements of
$\widehat{V}_{NN}$ between symmetrized Hyperspherical Harmonics.}
\label{Fig:PotEn_symm_L0}%
\end{center}
\end{figure}

\begin{figure}[!htb]
\begin{center}
\includegraphics[width=0.85\linewidth]{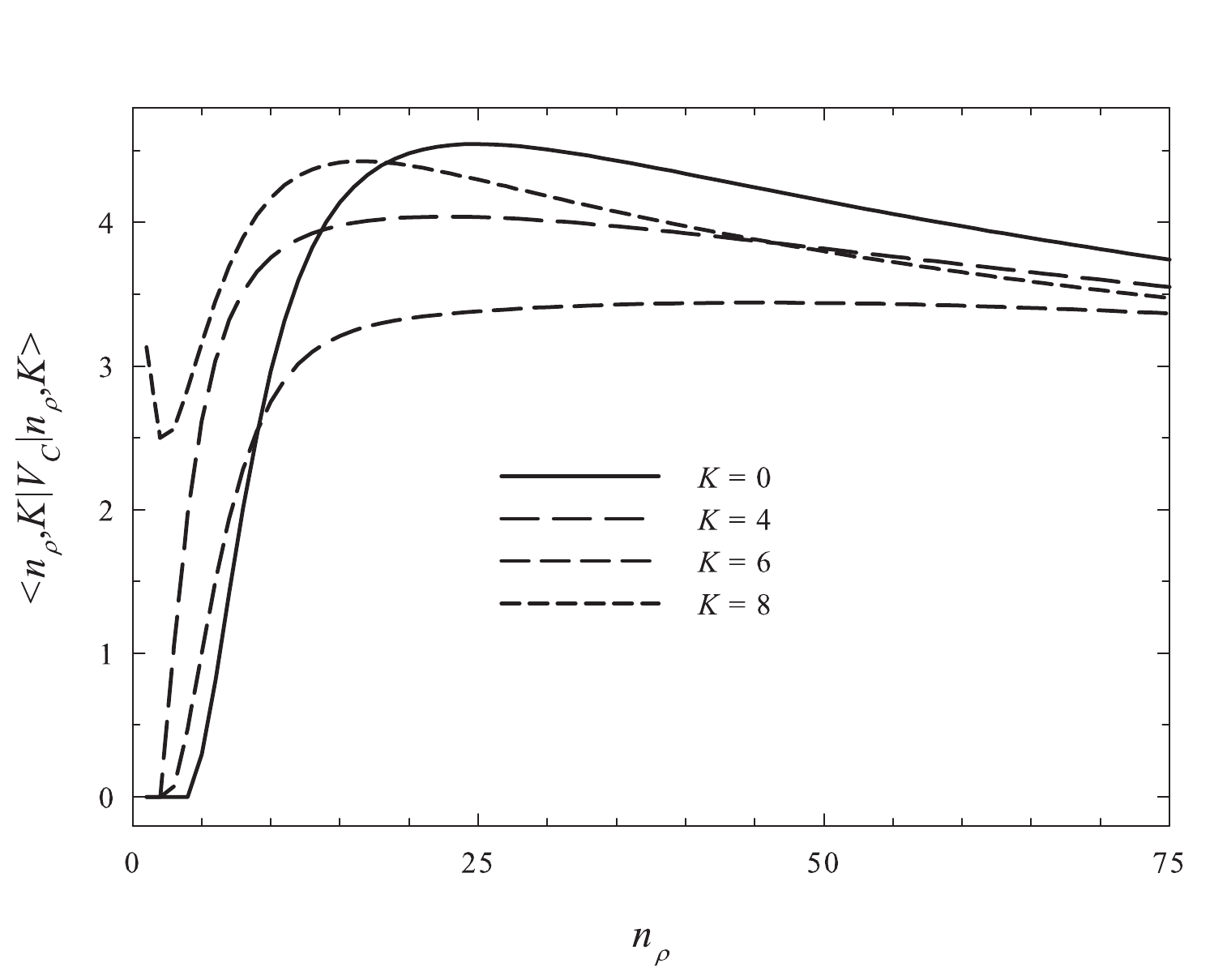}
\caption{Diagonal matrix elements of $\widehat{V}_{C}$
between symmetrized Hyperspherical Harmonics.}
\label{Fig:CoulPe_symm_L0}
\end{center}
\end{figure}

In Figures \ref{Fig:PotEn_symm_L0} and
\ref{Fig:CoulPe_symm_L0} the diagonal matrix elements of the nucleon-nucleon and Coulomb interactions
within the AMHHB model are displayed, again for channels up to $K=8$.
One observes that the nucleon-nucleon interaction creates a deep potential well
with a long tail in the Hyperspherical coordinate. This tail reflects the asymptotic form of the
potential, indicated in (\ref{eq:NN_asymp}). The matrix elements of the Coulomb interaction
indicate the magnitude of the Coulomb barrier, which is the main factor for generating the resonance states in $^{12}C$.


\subsection{Phase shifts and eigenphases}

\begin{figure}[ht!]
\begin{center}
\includegraphics[width=0.85\linewidth]{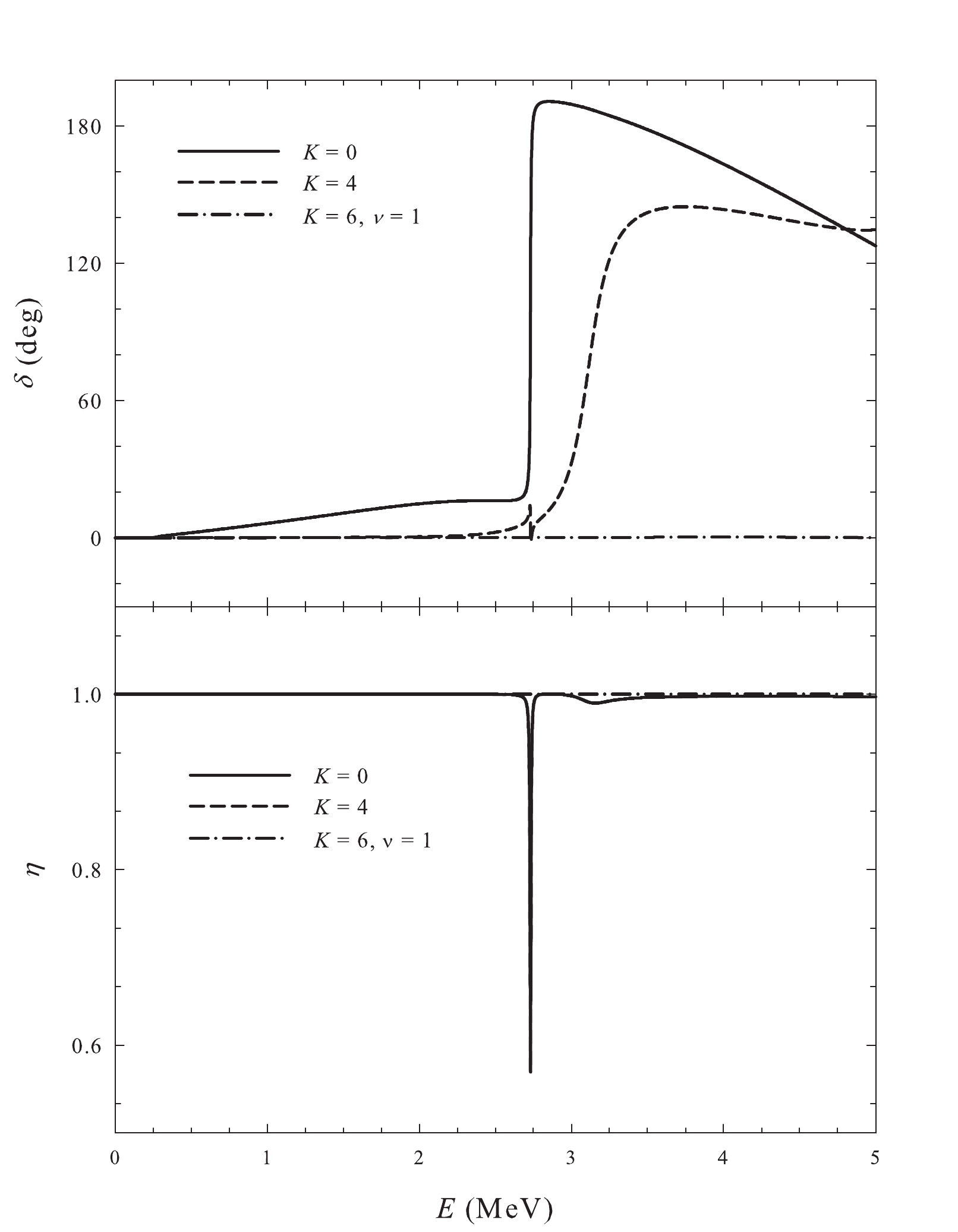}
\caption{Diagonal phase shifts and inelastic parameters for the $J^{\pi}=2^{+}$ state.}
\label{Fig:PhasesEtasL2}
\end{center}
\end{figure}

In Figure \ref{Fig:PhasesEtasL2} we show results of the AMHHB calculations
for the $2^{+}$ state in terms of the symmetrical Hyperspherical Harmonic
channels through
the (diagonal) phase shifts $\delta_{ii}$ and the inelastic
parameters $\eta_{ii}$ .

The scattering parameters are obtained from a calculation with maximal Hypermomentum
$K_{\max}=14$. One observes from Figure \ref{Fig:PhasesEtasL2}
that for small energies the channels are totally uncoupled ($\eta_{ii}\approx1$).
A first $2^{+}$ resonance appears at $E=2.731$ MeV, and is mainly produced in
the first channel with
Hypermomentum $K=2$, whereas a second
resonance at energy $E=3.113$ MeV is dominated by Hypermomentum $K=4$.
The inelastic
parameters for the first two channels have a pronounced minimum at the energy of the first resonance, and
a shallow minimum at the second resonance energy.
Also, the
first resonance displays a ``shadow resonance'' behavior in the second channel.
This is a typical behavior for resonances in a many-channel system (see,
for instance, the detailed analysis of two-channel resonances in $^{5}He$ in
\cite{kn:wilderm_eng}). The minimum in the inelastic parameters indicates that the
compound system is being reconstructed at this energy, and transits from one
channel to another.%

\begin{figure}[!ht]
\begin{center}
\includegraphics[width=0.85\linewidth]{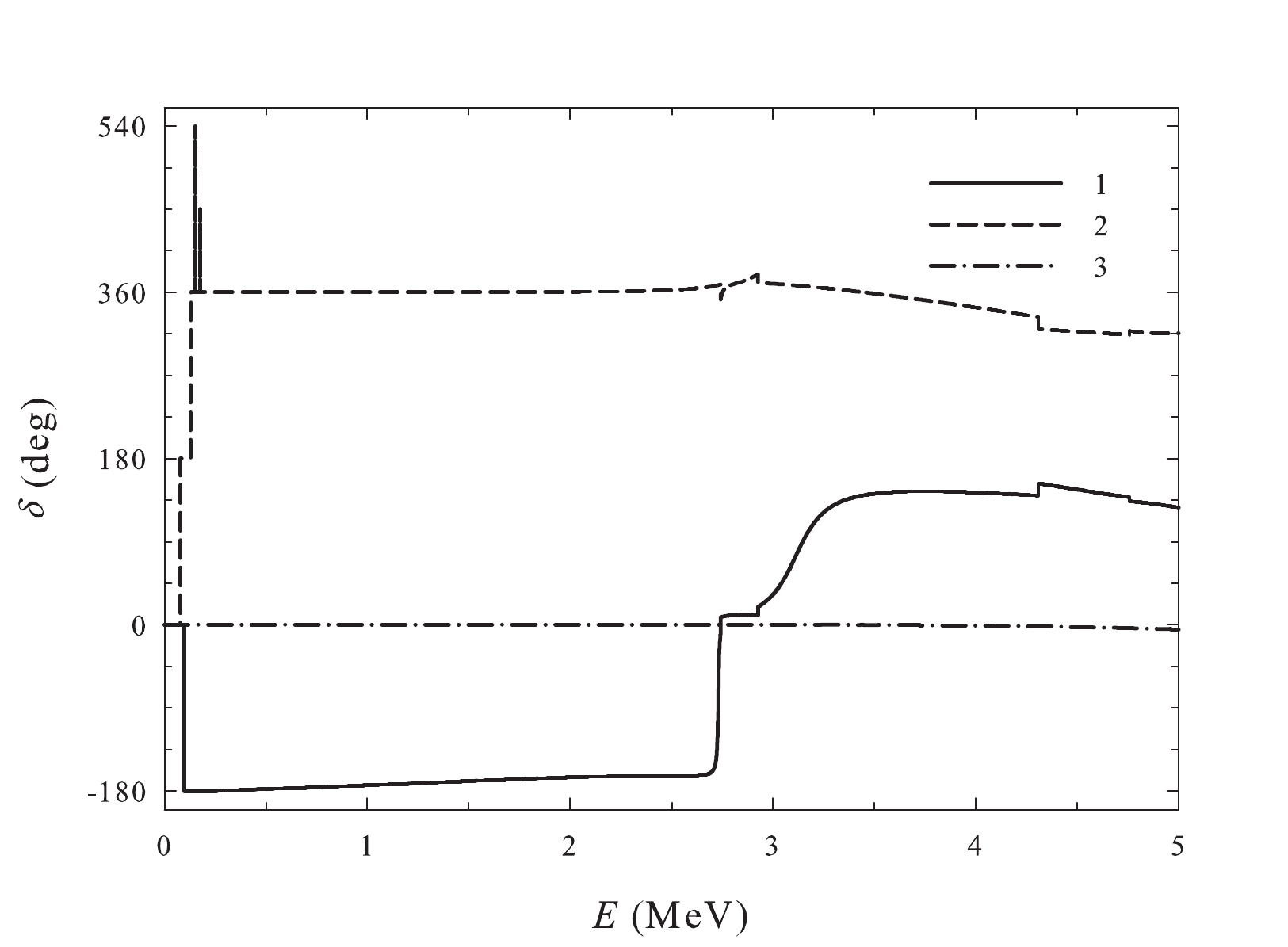}
\caption{Eigenphase shifts for $J^{\pi}=2^{+}$ for the first three eigenchannels.}
\label{Fig:EigenPhasL2}
\end{center}
\end{figure}

In Figure \ref{Fig:EigenPhasL2} we display the corresponding eigenphase shifts $\delta_{\alpha}$ for
the first three eigenchannels. One observes now that both resonance states are mainly associated
with the first eigenchannel, and that the second eigenchannel only contributes marginally.

\subsection{Convergence properties}

A convergence study of the
energies (and widths) for bound and resonance states should indicate whether the
Hilbert space is sufficiently large for stable and reliable results. The
AMHHB model space is characterized by two parameters:
the maximal value of Hypermomentum $K_{\max}$, and the maximal value of
the Hyperradial excitation $n_{\rho_{\max}}$. Usually the choice is a compromise
between the convergence of the results and the computational burden. A set of
Hyperspherical Harmonics with $K_{\max}=14$ for even parity states, and
$K_{\max}= 13$ for odd parity states, seems sufficient and remains computationally feasible.
This choice accounts
for a large number of three-cluster configurations or, in other words, for a sufficient
number of inherent (triangular) shapes for the three clusters. We refer to
\cite{2001PhRvC..63f4604V} for examples of most probable triangular shapes 
for the Hyperspherical Harmonics from $K=0$ to $K=10$.

\begin{figure}[!ht]
\begin{center}
\includegraphics[width=0.85\linewidth]{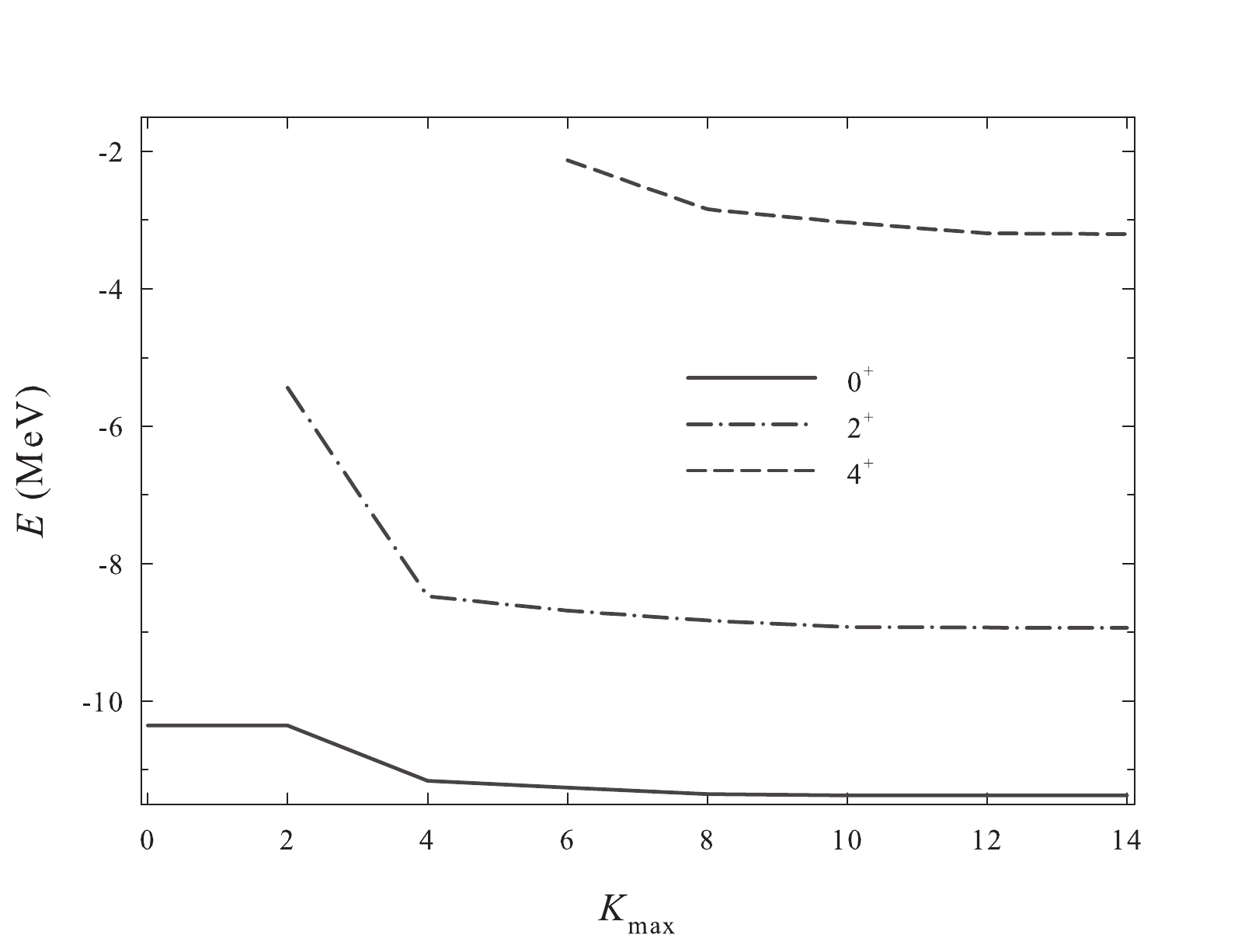}
\caption{Convergence of the bound states in AMHHB.}
\label{Fig:SpectrConvN}
\end{center}
\end{figure}

A first convergence test considers the $0^{+}$, $2^{+}$ and $4^{+}$ bound states of $^{12}C$, shown in Figure \ref{Fig:SpectrConvN}
as a function of $K_{\max}$.
One observes that the deeply bound states ($J^{\pi}=0^{+}$,
$2^{+}$) require significantly less Hyperspherical Harmonics for a converged
energy than the shallow, or weakly bound, state with $J^{\pi}=4^{+}$. At least all
Hyperspherical Harmonics with $K_{\max}\geq6$ are required to bind
the latter state, whereas for  $J^{\pi}=0^{+}$ one already obtains binding
with a single 
Hyperspherical Harmonic with $K=0$. Figure \ref{Fig:SpectrConvN} further demonstrates
that the above choice of $K_{\max}$ amply leads to sufficient precision for the bound states.

In Table \ref{Tab:Resonance0_K=12MArai} we turn to the energies and widths of the
$0^{+}$ and $2^{+}$ resonances obtained with increasing number of
Hyperspherical Harmonics. One observes that sufficient convergence of the resonances 
occurs at $K_{\max}=12$. It is furthermore interesting to note that these resonances
already appear with reasonable energy and width values when only the lowest channel
($K=0$ for the $0^+$, and $K=2$ for the $2^+$ state) is considered. This is a remarkable
result for $^{12}C$, as e.g.\ for $^{6}Be$ it was impossible to generate a $0^{+}$
resonance with a single $K=0$ channel (see \cite{2001PhRvC..63c4607V}).

\begin{table}[!ht] \centering
\caption{Energy (MeV) and width (keV) of the low-lying resonances in terms of
$K_{\max}$}%
\begin{tabular}
[c]{|c|c|r|r|r|r|r|r|r|}\hline
$L^{\pi}$ & $K_{\max}$ & 0 & 4 & 6 & 8 & 10 & 12 & 14 \\\hline
$0^{+}$ & $E$ & 0.40 & 0.75 & 0.74 & 0.72 & 0.70 & 0.68 & 0.68 \\
& $\Gamma$ & 205.08 & 13.40 & 11.79 & 7.10 & 4.35 & 2.71 & 2.77 \\\hline
$0^{+}$ & $E$ & 1.15 & 7.34 & 6.09 & 5.55 & 5.54 & 5.16 & 5.14 \\
& $\Gamma$ & 510.16 & 897.64 & 422.50 & 539.21 & 586.08 & 534.33 & 523.46 \\\hline
$2^{+}$ & $E$ & - & 3.28 & 2.89 & 2.83 & 2.78 & 2.74 & 2.73 \\
& $\Gamma$ & - & 30.19 & 13.07 & 11.85 & 9.95 & 8.84 & 8.75 \\\hline
$2^{+}$ & $E$ & - & 3.50 & 3.27 & 3.22 & 3.17 & 3.14 & 3.11 \\
& $\Gamma$ & - & 274.51 & 351.57 & 308.29 & 280.23 & 263.80 & 246.78 \\\hline
\end{tabular}
\label{Tab:Resonance0_K=12MArai}%
\end{table}%

In all calculations we have considered states with hyperradial excitation up to $n_{\rho \max}$=70, which covers a
large range of intercluster distances, and reaches well into the asymptotic region.

\subsection{Partial widths.}

In Table \ref{Tab:PartWidthsE} we display the energy, the
total width ($\Gamma$) and the partial widths ($\Gamma_{i}$, $i=1,2,\ldots$) in the
corresponding decay channels for the even parity resonances, and in
Table \ref{Tab:PartWidthsO} for the odd parity resonances.
\begin{table}[!ht] \centering
\caption{Partial widths of the even parity resonances in $^{12}C$. Energy in MeV,
widths in keV.}%
\begin{tabular}
[c]{|c|c|r|c|r|c|r|c|r|}\hline
$L^{\pi}$ & \multicolumn{2}{|c}{$0^{+}$} & \multicolumn{2}{|c}{$2^{+}$} &
\multicolumn{2}{|c}{$2^{+}$} & \multicolumn{2}{|c|}{$4^{+}$} \\\hline
$E$ & \multicolumn{2}{|r|}{0.68} & \multicolumn{2}{|r|}{2.78} &
\multicolumn{2}{|r|}{3.17} & \multicolumn{2}{|r|}{5.60} \\\hline
$\Gamma$ & \multicolumn{2}{|r|}{2.79} & \multicolumn{2}{|r|}{9.95} &
\multicolumn{2}{|r|}{280.24} & \multicolumn{2}{|r|}{0.55} \\\hline
$\Gamma_{1}$ & $K=0$ & 2.79& $K=2$ & 6.11& $K=2$ & 13.46 & $K=4$ & 0.23\\\hline
$\Gamma_{2}$ & $K=4$  & 0 & $K=4$ & 3.84& $K=4$ & 278.89 & $K=6$ & 0.15 \\\hline
$\Gamma_{3}$ & $K=6$ & 0 & $K=6$ & $<$10$^{-5}$& $K=6$ & $<$10$^{-5}$& $K=8$ & 0.16\\\hline
\end{tabular}
\label{Tab:PartWidthsE}
\end{table}

\begin{table}[!ht] \centering
\caption{Partial widths of the odd parity resonances in $^{12}C$. Energy in MeV,
widths in keV.}%
\begin{tabular}
[c]{|c|c|r|c|r|}\hline
$L^{\pi}$ & \multicolumn{2}{|c}{$1^{-}$} & \multicolumn{2}{|c|}{$3^{-}$}\\\hline
$E$ & \multicolumn{2}{|r|}{3.52} & \multicolumn{2}{|r|}{0.67}\\\hline
$\Gamma$ &\multicolumn{2}{|r|}{0.21} & \multicolumn{2}{|r|}{8.34}\\\hline
$\Gamma_{1}$  & $K=3$& 0.206 & $K=3$&8.34\\\hline
$\Gamma_{2}$& $K=5$  &0.002& $K=5$& 0\\\hline
$\Gamma_{3}$ & $K=7$& $<$10$^{-5}$& $K=7 $& 0\\\hline
\end{tabular}
\label{Tab:PartWidthsO}
\end{table}

One observes that in most cases only one or two channels are responsible for the decay of the resonance
states. The remaining channels contribute negligibly, and the corresponding partial width
does not exceed $10^{-5}$ keV. Only for the $4^+$ resonance a significant distribution over multiple
channels is apparent.

One should note that,
although the resonances are created by only a few channels, the role of
the other, very weakly coupled, channels is still important. This can be seen from Table
\ref{Tab:Resonance0_K=12MArai} for the first $0^{+}$ resonance: it is indeed generated
mainly by the channel with minimal Hypermomentum $K=0$, but modified substantially with
increasing number of Hypermomentum. The same applies to the other resonance states.

\subsection{Correlation functions and density distributions.}

\begin{figure}[!ht]
\begin{center}
\includegraphics[width=0.85\linewidth]{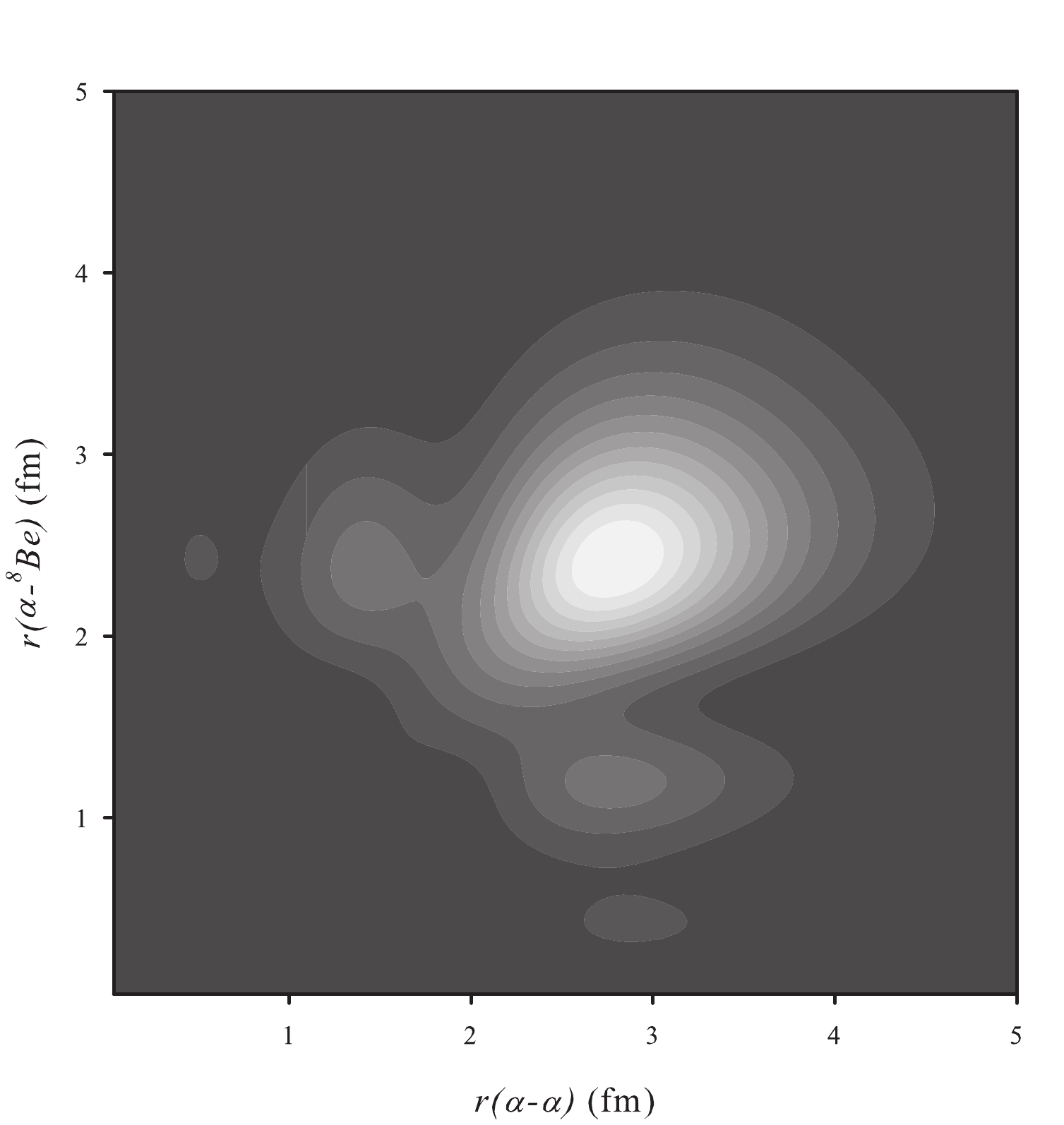}
\caption{Correlation function for the $^{12}C$ ground state in coordinate space.}
\label{Fig:CorrFunBS_CS_L0}
\end{center}
\end{figure}

In Figure \ref{Fig:CorrFunBS_CS_L0} we show the correlation function for the
$^{12}C$ ground state, and observe that this state displays a compact
spatial configuration, as it is expected for such a deeply bound state. The most
probable shape of the three $\alpha$-cluster system is an almost
equilateral triangle with a distance between any two $\alpha$-particles of approximately
3 fm.
\begin{figure}[!ht]
\begin{center}
\includegraphics[width=0.85\linewidth]{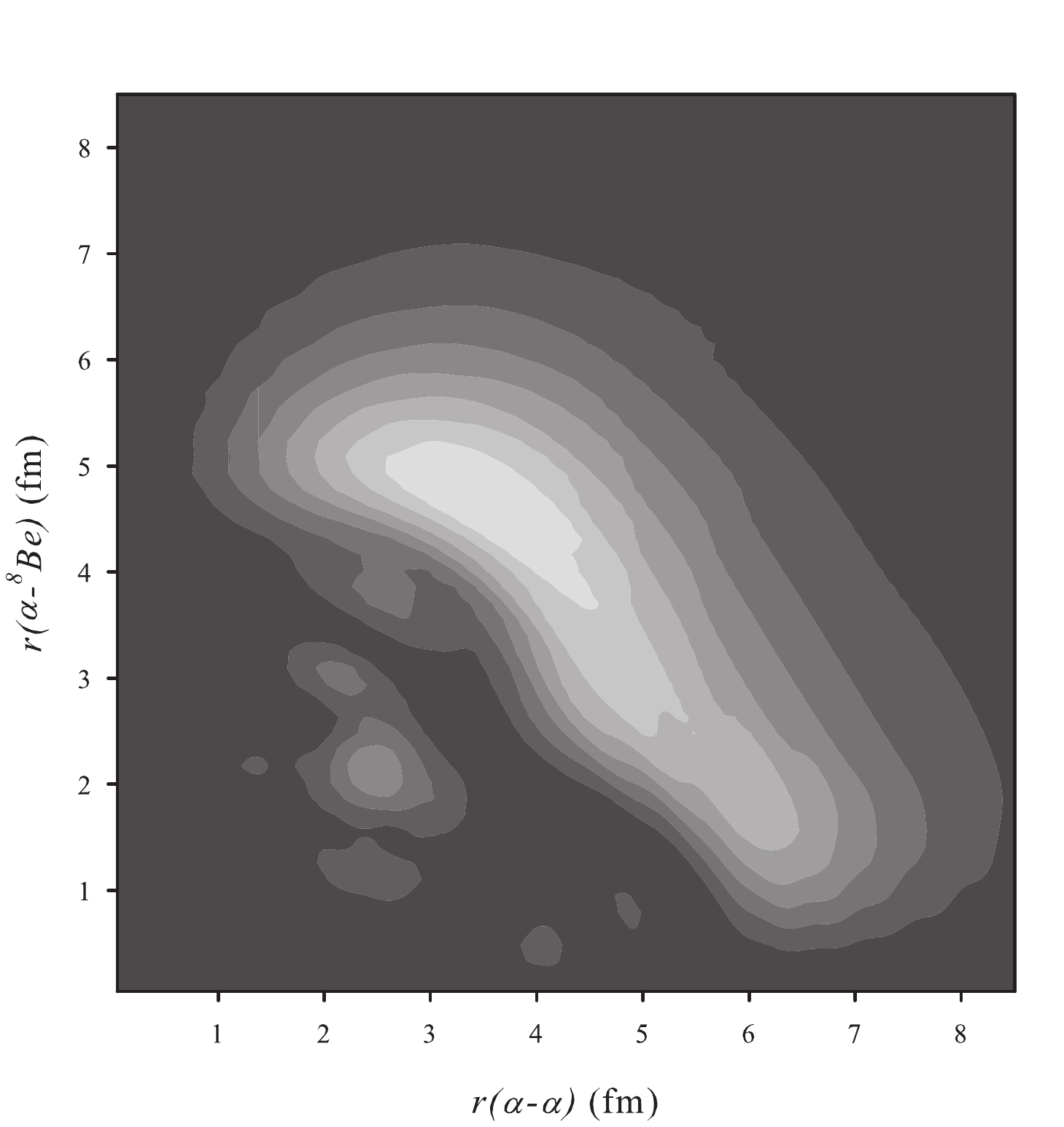}
\caption{Correlation function for the first $0^+$ resonance state of $^{12}C$ in coordinate space.}
\label{Fig:CorrFun_CS_R1_L0}
\end{center}
\end{figure}

The correlation function for the first $0^+$ resonance
state on the other hand, shown in Figure \ref{Fig:CorrFun_CS_R1_L0}, shows a more deformed system
with two $\alpha$ particles relatively close to one another
(about 3.5 fm) and the third alpha-particle further away (approximately 5 fm). So $^{12}C$ features
a prolate triangle as a dominant configuration for this state.

One also observes on Figure \ref{Fig:CorrFun_CS_R1_L0} a small maximum for the correlation function
corresponding to an almost linear configuration of three $\alpha$ particles, two of them being approximately
4 fm apart, and the third 0.2 fm away from their centre of mass. However, the weight of this linear
configuration is approximately 6 times less than the weight of the prolate triangular configuration.
Our calculations therefore do not agree with other authors advancing a dominant linear structure
\cite{1995PhRvC..51..152A,1992NuPhA.549..431M,2007PThPh.117..655K,2004NuPhA.738..357N}.  
\begin{figure}[!ht]
\begin{center}
\includegraphics[width=0.85\linewidth]{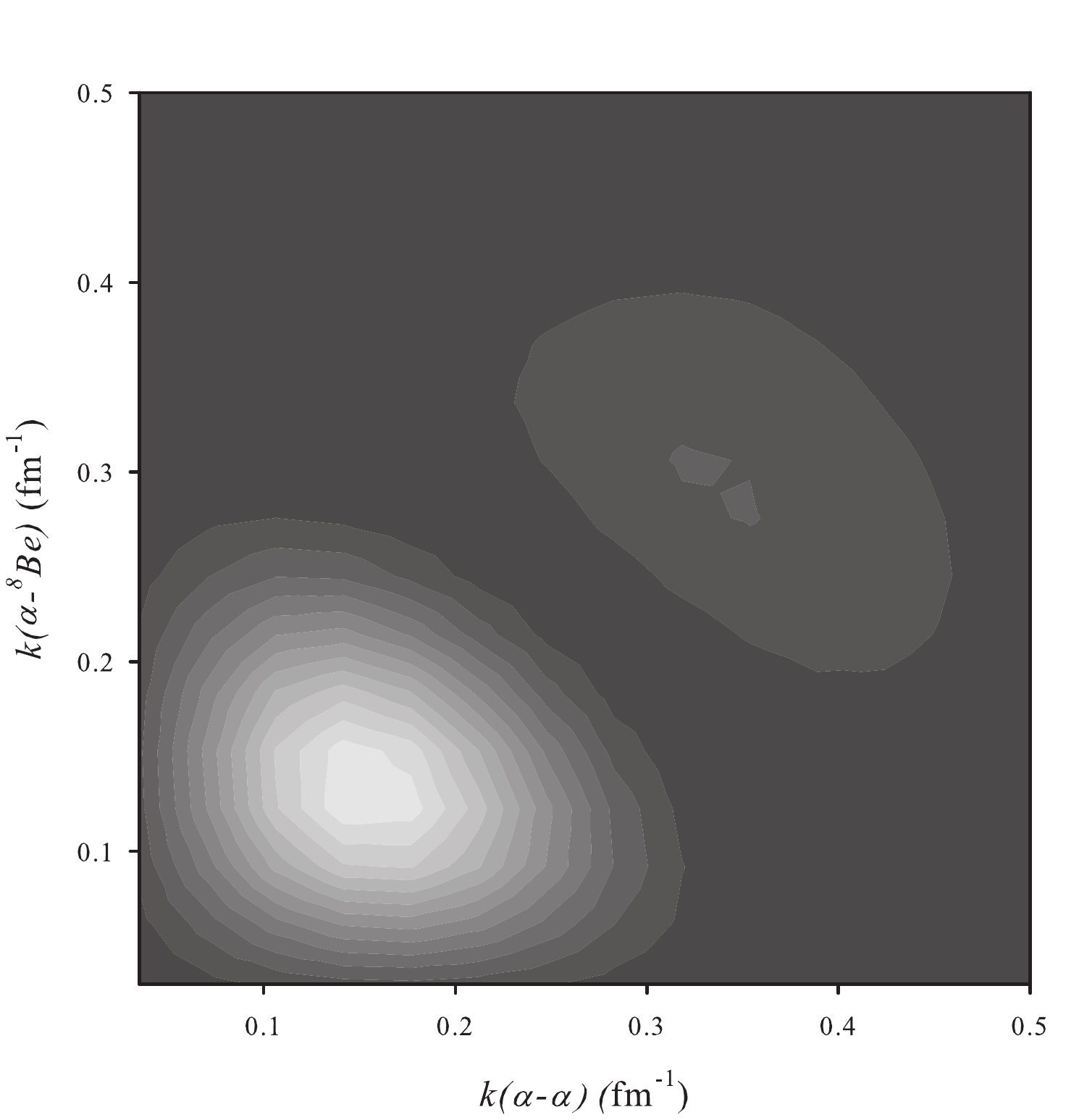}
\caption{Correlation function for the first $0^+$ resonance state of $^{12}C$ in momentum space.}
\label{Fig:CorrFun_MS_R1_L0}
\end{center}
\end{figure}

In Figure \ref{Fig:CorrFun_MS_R1_L0} we display the correlation
function of the first resonance state in momentum space. One observes a huge
maximum corresponding to relatively slowly moving $\alpha$ particles. A
small maximum corresponding to faster moving alpha-particles is also present.

\subsection{Comparison to the literature}

We now compare the AMHHB results to the existing literature. In
Table \ref{Tab:AMvsCSMAria} we display the AMHHB results to those of Arai
\cite{2006PhRvC..74f4311A} and Pichler et al.\
\cite{1997NuPhA.618...55P}, both obtained by the Complex Scaling Method (CSM). 
The latter authors \cite{1997NuPhA.618...55P} use a somewhat different value for the parameter $u$ in the Minnesota potential, and a different oscillator length $b$; because of this, different results are obtained for the bound states.
\begin{table}[!ht] \centering
\caption{Bound and resonance states of $^{12}C$ obtained with the AMHHB model, compared to CSM results
from the literature.}
\begin{tabular}
[c]{|c|r|r|r|r|r|r|}\hline
Method & \multicolumn{2}{|c|}{AMHHB} & \multicolumn{2}{|c|}{CSM-Arai} &
\multicolumn{2}{|c|}{CSM-Pichler et al.}\\\hline
Reference & \multicolumn{2}{|c}{Present paper} &
\multicolumn{2}{|c}{\cite{2006PhRvC..74f4311A}} &
\multicolumn{2}{|c|}{\cite{1997NuPhA.618...55P}}\\\hline
$J^{\pi}$ & $E$, MeV & $\Gamma$, keV & $E$, MeV & $\Gamma$, keV & $E$, MeV &
$\Gamma$, keV\\\hline
$0^{+}$ & $-11.372$ & & $-11.37$ & & $-10.43$ & \\
& 0.684 & 2.71 & 0.4 &
$<1$ & 0.64 & 14\\
& 5.156 & 534.00 & 4.7 & 1000 & 5.43 & 920\\\hline
$2^{+}$ & $-8.931$ & & $-8.93$ &  & $-7.63$ & \\
& 2.775 & 9.95 & 2.1 & 800 & 6.39 & 1100\\
& 3.170 & 280.24 & 4.9 & 900 &  & \\\hline
$4^{+}$ & $-3.208$ & & $-3.21$ &  &  & \\
& 5.603 & 7.82 & 5.1 & 2000 &  & \\ \hline
$1^{-}$ & 3.516 & 0.21 & 3.4 & 200 & 3.71 & 360\\\hline
$3^{-}$ & 0.672 & 8.34 & 0.6 &
$<50$ & 1.16 & 25\\
& 4.348 & 2.89 & 7.1 & 5400 & 11.91 & 1690\\
& 5.433 & 334.90 & 9.6 & 400 &  & \\\hline
\end{tabular}
\label{Tab:AMvsCSMAria}\end{table}

Comparison with the results of Arai \cite{2006PhRvC..74f4311A} indicates that
the AMHHB model leads to resonance states with higher
energy and smaller widths than those obtained with the CSM.
This can be attributed to the difference in the methods, and to the different Hilbert spaces.
Formally the Hilbert space of basis functions
used by Arai \cite{2006PhRvC..74f4311A} is quite close to the one considered in the AMHHB.
Actually, in the present calculations the partial orbital
momenta $l_{1}$ and $l_{2}$ are restricted by the condition%
\[
L\leq l_{1}+l_{2}\leq K_{\max}%
\]
so that, for instance, for total orbital momentum $L=0$, they run from
$l_{1}=l_{2}=0$ to  $l_{1}=l_{2}=6$ with $K_{\max}=14$. Arai on the other hand, restricted
himself with $l_{1},l_{2}\leq4$. In \cite{2001PhRvC..63c4607V,2007JPhG...34.1955B,A4Resonances2004}
we observed the tendency  that
the more Hyperspherical Harmonics (thus the more channels) are involved in
the calculation, the smaller the resonance energy and width becomes. This
tendency is again confirmed by the present AMHHB calculations. Thus some reduction of
the width of the resonances, observed in our calculations with respect to Arai \cite{2006PhRvC..74f4311A},
can be attributed to
the larger number of channels in our model.

Comparing the AMHHB results to the
Complex Scaling Model calculations of Pichler et al.\ \cite{1997NuPhA.618...55P}, one
observes that both yield
close results for the first and second  $0^{+}$ resonance states.

On the whole one can conclude that there is consistency in the results for resonance properties
in all three microscopic models. 

\subsection{Comparison to experiment}

In Table \ref{Tab:TheoryAndExperiment} we compare the theoretical AMHHB results for $^{12}C$
to available experimental data.

\begin{table}[!t] \centering
\caption{Bound and resonance states of $^{12}C$ obtained with the
AMHHB model, compared to experiment. }%
\begin{tabular}
[c]{|c|r|r|r|r|}\hline
Method & \multicolumn{2}{|c|}{AMHHB} & \multicolumn{2}{|c|}{Experiment } \\
\hline
Reference & \multicolumn{2}{|c}{Present paper} & \multicolumn{2}{|c|}{\cite{1990NuPhA.506....1A}} \\
\hline
$J^{\pi}$ & $E$, MeV & $\Gamma$, keV & \multicolumn{1}{|c|}{$E$, MeV} & \multicolumn{1}{|c|}{$\Gamma$, keV} \\
\hline
$0^{+}$ & $-11.372$ &  & $-7.2746$ &  \\
  & 0.684 & 2.71 & $0.3796 \pm 0.0002$ & $(8.5 \pm 1.0) \times10^{-3}$ \\
  & 5.156 & 534.00 & $3.0 \pm 0.3$ & $3000 \pm 700$ \\
\hline
$2^{+}$ & $-8.931$ &  & $-2.8357 \pm 0.0003$ &  \\
  & 2.775 & 9.95 & $3.89 \pm 0.05$ & $430 \pm 80$ \\
  & 3.170 & 280.24 & $8.17 \pm 0.04$ & $1500 \pm 200$ \\
\hline
$4^{+}$ & $-3.208$ &  &  & \\
  & 5.603 & 7.82 & $6.808 \pm 0.015$ & $258 \pm 15$ \\
\hline
$1^{-}$ & 3.516 & 0.21 & $3.569 \pm 0.016$ & $315 \pm 25$ \\ \hline
$3^{-}$ & 0.672 & 8.34 & $2.366 \pm 0.005$ & $34 \pm 5$ \\
& 4.348 & 2.89 &  & \\
& 5.433 & 334.90 &  & \\\hline
\end{tabular}
\label{Tab:TheoryAndExperiment}%
\end{table}

One notices that the first $0^{+}$ resonance state (the Hoyle state) appears in the current
calculations  as a narrow resonance with an energy of 0.684 MeV and width 2.7 keV,
which is considerably wider than the experimental Hoyle state (about 8.5 $10^{-3}$ keV).
This contrasts with the generally observed
feature of the AMHHB calculations that the calculated widths
are significantly less than the corresponding experimental widths of the $^{12}C$ resonances.
The discrepancies between the theoretical and experimental data have essentially two origins.
The first one relates to the choice of the nucleon-nucleon interaction: it has been 
tuned to reproduce the phase shifts and resonance properties for alpha-alpha
scattering. As a result it leads to overbound $0^{+}$ and $2^{+}$ states in $^{12}C$, and binds 
the $4^{+}$ state. The second one relates to the specific choice of three-cluster model and corresponding
model space, as well as to the method by which the energy and width of the resonance states are obtained.

\subsection{Optimizing the nucleon-nucleon potential}

In this paper we used a Minnesota nucleon-nucleon potential
tuned to reproduce the phase shifts for $\alpha-\alpha$ scattering, as well as the
$^{8}Be$ resonances. This however leads to overbound $0^+$ and $2^+$ states,
and a bound $4^+$ state. Moreover, the obtained resonance structure for the $^{12}C$ three-cluster
continuum deviates from the experimentally observed one, which can also be attributed
to the specific choice of semi-realistic nucleon-nucleon potential.

We therefore wish to discuss the dependence of the results to the choice of parameter $u$
on the results. To do so we use different criteria to optimize this parameter.
We first determine a value to reproduce the ground state energy of $^{12}C$, followed
by an attempt to reproduce the energy and width of the $0^+$ Hoyle state.

\begin{figure}[!ht]
	\begin{center}
			\includegraphics[width=0.85\linewidth]{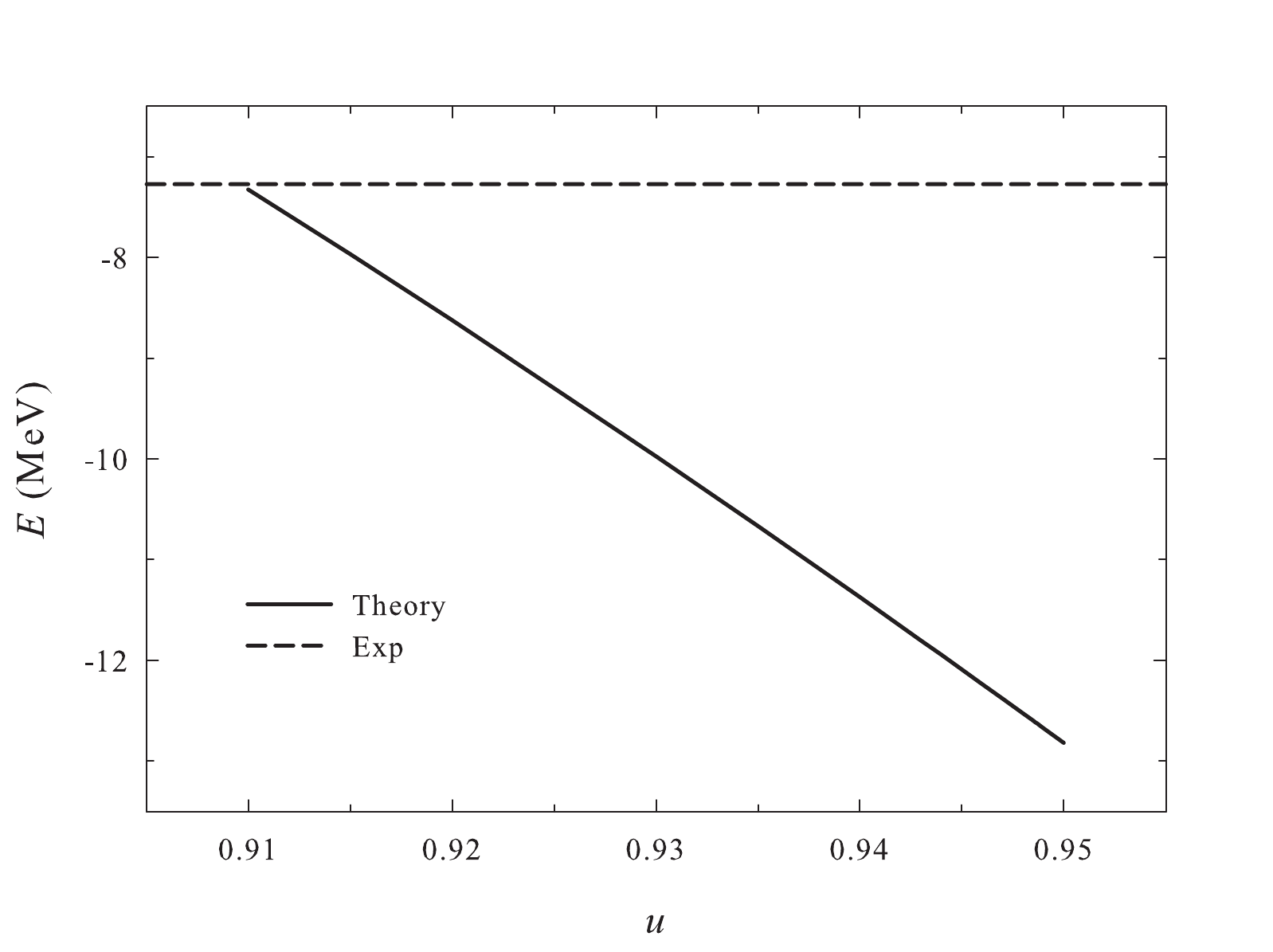}
			\caption{Energy of the ground state as a function of parameter $u$ of the Minnesota potential.}
			\label{Fig:Spectr_L0_VU}
		\end{center}
\end{figure}

\begin{figure}[!ht]
	\begin{center}
		\includegraphics[width=0.85\linewidth]{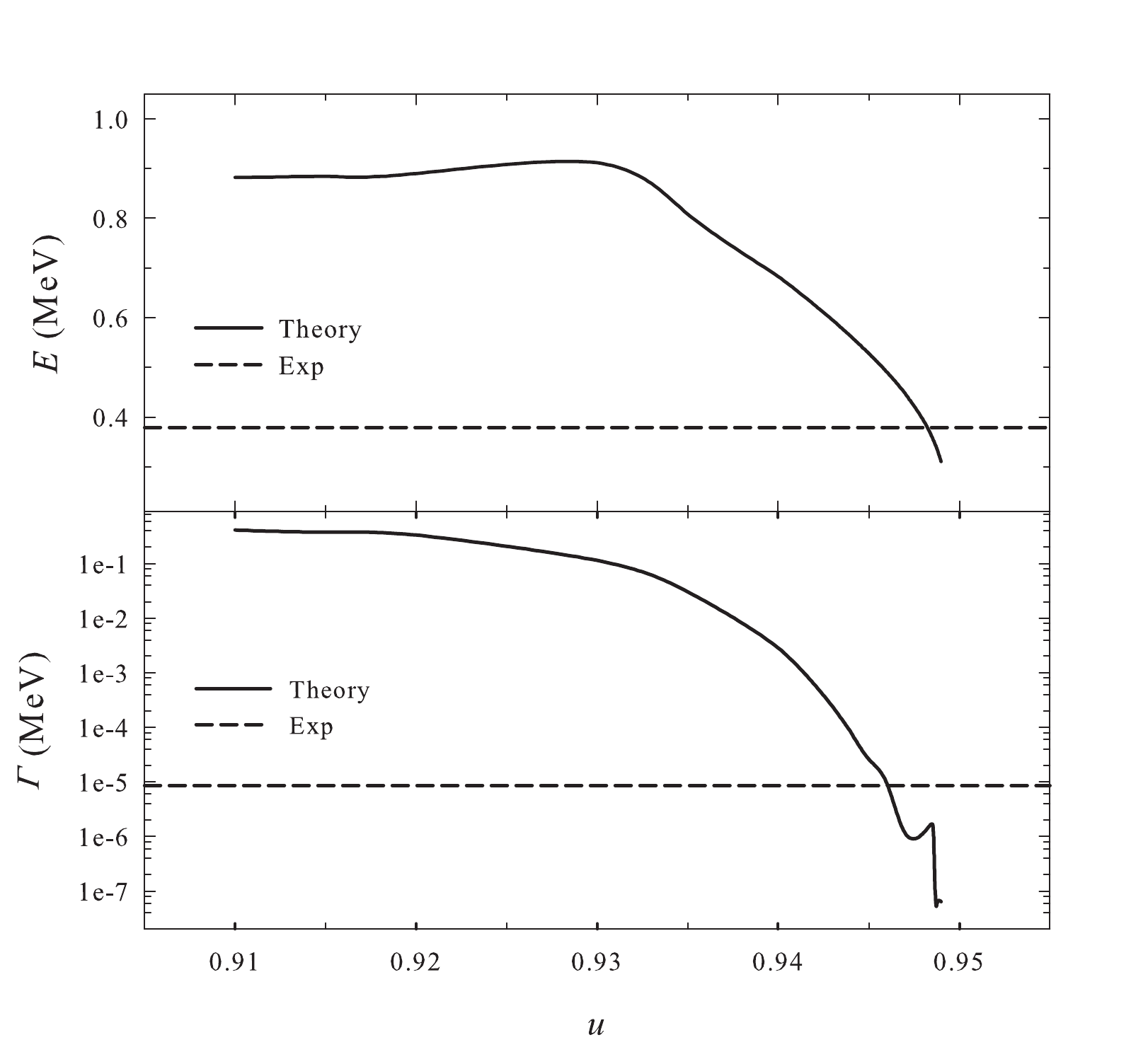}
		\caption{Position and total width of the first $0^{+}$ resonance state as a function of parameter $u$.}
		\label{Fig:Gamma_L0_VU}
	\end{center}
\end{figure}

In Figure \ref{Fig:Spectr_L0_VU} we display the ground state energy as a function of the parameter $u$,
compared to experiment (dashed line). One observes that the ground state is reproduced with $u=0.910$.
One observes a monotonously decreasing linear dependence of the ground state energy on $u$ within the selected range.
For the Hoyle state position and width the dependency is less trivial, as is shown in Figure \ref{Fig:Gamma_L0_VU}. One
however observes
that 
the value $u=0.948$ reproduces the position of the Hoyle state, and leads to a close match for its width too.

The correlation functions for the ground state and Hoyle state obtained with their
respective optimal values were very close to the ones obtained with the value $u=0.94$ and 
displayed in Figures \ref{Fig:CorrFunBS_CS_L0} and\ \ref{Fig:CorrFun_CS_R1_L0}, so that the
conclusions remain unaltered.

\section{Conclusions}

In this paper we described the $^{12}C$ nucleus with a three-cluster microscopic
model.

The
model correctly handles the three-cluster continuum, i.e.\ correctly implements the suitable boundary conditions, by using a
Hyperspherical Harmonics basis. It leads to the scattering matrix $S$ in many-channel space, and the energy, total and
partial widths of the resonance states and their corresponding wave functions can be obtained.

It was shown that the obtained resonances of $^{12}C$ agree
well with other methods, and that the lowest resonances are generated by only a few number of
weakly coupled channels, leading
narrow resonance states. The partial widths determine the most
probable channels for resonance decay. Correlation functions and
density distributions revealed the dominant shape of the three-cluster triangle configuration
for the lowest bound and resonance states of $^{12}C$. There were no indications of
a prominent linear three-cluster structure for the resonance states.

It was also shown that it is impossible to fix a unique value for the $u$ parameter of the Minnesota nucleon-nucleon potential
to fit all desired physical properties for $^{12}C$, and for the disintegrating $\alpha$ particles. However the
qualitative conclusions remained unaltered under slight adaptation of $u$.

As a final conclusion we can state that the model is  consistent with other microscopic
models using the Complex Scaling methodology to determine the energy and
total width of three-cluster resonance states.

\section{Acknowledgments}

Support from the Fonds voor Wetenschappelijk Onderzoek Vlaanderen (FWO),
G0120-08N is gratefully acknowledged. V. S. Vasilevsky is grateful to the
Department of Mathematics and Computer Science of the University of Antwerp
(UA) for hospitality. This work was supported in part by the Program of
Fundamental Research of the Physics and Astronomy Department of the National
Academy of Sciences of Ukraine.

\bibliographystyle{ieeetr}
\bibliography{12C,Aoyama,Arai,BayeDescouvemont,CMPLX_SC,Csoto,Experiment,Fedorov,Filippov,Fujiwara,Kato,My_publE,Suzuki,THESIS,THREE_CL}

\begin{thebibliography}{10}

\bibitem{1990NuPhA.506....1A}
F.~{Ajzenberg-Selove}, ``{Energy levels of light nuclei \makebox{$A$} =
  11-12},'' {\em Nucl. Phys. A}, vol.~506, pp.~1--158, Jan. 1990.

\bibitem{2004PhRvC..70a4006F}
S.~I. {Fedotov}, O.~I. {Kartavtsev}, V.~I. {Kochkin}, and A.~V. {Malykh},
  ``{3{$\alpha$} -cluster structure of the {$0^{+}$} states in {$^{12}C$} and
  the effective {$\alpha$}-{$\alpha$} interactions},'' {\em Phys. Rev. C},
  vol.~70, p.~014006, July 2004.

\bibitem{2005JPhG...31.1207F}
I.~{Filikhin}, V.~M. {Suslov}, and B.~{Vlahovic}, ``{{$0^{+}$} states of the
  {$^{12}C$} nucleus: the Faddeev calculation in configuration space},'' {\em
  Journal of Physics G Nuclear Physics}, vol.~31, pp.~1207--1224, Nov. 2005.

\bibitem{2007EPJA...31..303A}
R.~{{\'A}lvarez-Rodr{\i}guez}, E.~{Garrido}, A.~S. {Jensen}, D.~V. {Fedorov},
  and H.~O.~U. {Fynbo}, ``{Structure of low-lying {$^{12}C$} resonances},''
  {\em Eur. Phys. J. A}, vol.~31, pp.~303--317, Mar. 2007.

\bibitem{2008JPhCS.111a2017A}
R.~{{\'A}lvarez-Rodr{\'{\i}}guez}, A.~S. {Jensen}, D.~V. {Fedorov}, H.~O.~U.
  {Fynbo}, and E.~{Garrido}, ``{Decay of low-lying {$^{12}C$} resonances within
  a 3{$\alpha$} cluster model},'' {\em J. Phys. Conf. Ser.}, vol.~111,
  p.~012017, May 2008.

\bibitem{2008PhRvC..77f4305A}
R.~{{\'A}lvarez-Rodr{\'{\i}}guez}, A.~S. {Jensen}, E.~{Garrido}, D.~V.
  {Fedorov}, and H.~O.~U. {Fynbo}, ``{Momentum distributions of {$\alpha$}
  particles from decaying low-lying {$^{12}C$} resonances},'' {\em Phys. Rev.
  C}, vol.~77, p.~064305, June 2008.

\bibitem{2010PhRvL.105b2501C}
M.~{Chernykh}, H.~{Feldmeier}, T.~{Neff}, P.~{von Neumann-Cosel}, and
  A.~{Richter}, ``{Pair Decay Width of the Hoyle State and its Role for Stellar
  Carbon Production},'' {\em Phys. Rev. Lett.}, vol.~105, p.~022501, July 2010.

\bibitem{2011PhLB..695..324D}
R.~{de Diego}, E.~{Garrido}, D.~V. {Fedorov}, and A.~S. {Jensen}, ``{The triple
  alpha reaction rate and the {$2^{+}$} resonances in {$^{12}C$}},'' {\em Phys.
  Lett. B}, vol.~695, pp.~324--328, Jan. 2011.

\bibitem{1997NuPhA.618...55P}
R.~{Pichler}, H.~{Oberhummer}, A.~{Cs{\' o}t{\' o}}, and S.~A. {Moszkowski},
  ``{Three-alpha structures in \makebox{$^{12}C$}},'' {\em Nucl. Phys. A},
  vol.~618, pp.~55--64, Feb. 1997.

\bibitem{2004NuPhA.738..455K}
C.~{Kurokawa} and K.~{Kat{\= o}}, ``{Three-alpha resonances in {$^{12}C$}},''
  {\em Nuclear Physics A}, vol.~738, pp.~455--458, June 2004.

\bibitem{2004NuPhA.738..495F}
Y.~{Fujiwara}, K.~{Miyagawa}, M.~{Kohno}, Y.~{Suzuki}, D.~{Baye}, and J.-M.
  {Sparenberg}, ``{A consistent 3{$\alpha$} and 2{$\alpha$}{$\Lambda$} Faddeev
  calculation using the 2{$\alpha$} RGM kernel},'' {\em Nucl. Phys. A},
  vol.~738, pp.~495--498, June 2004.

\bibitem{2004PhRvC..69c7002F}
Y.~{Fujiwara}, Y.~{Suzuki}, and M.~{Kohno}, ``{Case of almost redundant
  components in 3{$\alpha$} Faddeev equations},'' {\em Phys. Rev. C}, vol.~69,
  p.~037002, Mar. 2004.

\bibitem{2004FBS....34..237F}
Y.~{Fujiwara}, M.~{Kohno}, and Y.~{Suzuki}, ``{Solving Three-Cluster OCM
  Equations in the Faddeev Formalism},'' {\em Few-Body Systems}, vol.~34,
  pp.~237--257, 2004.

\bibitem{2002PThPh.107..745F}
Y.~{Fujiwara} and H.~{Nemura}, ``{Three-Cluster Equation Using Two-Cluster RGM
  Kernel},'' {\em Prog. Theor. Phys.}, vol.~107, pp.~745--757, Apr. 2002.

\bibitem{2004PhRvC..70b4002F}
Y.~{Fujiwara}, K.~{Miyagawa}, M.~{Kohno}, Y.~{Suzuki}, D.~{Baye}, and J.-M.
  {Sparenberg}, ``{Faddeev calculation of 3{$\alpha$} and
  {$\alpha$}{$\alpha$}{$\Lambda$} systems using {$\alpha$}{$\alpha$}
  resonating-group method kernels},'' {\em Phys. Rev. C}, vol.~70, p.~024002,
  Aug. 2004.

\bibitem{2005PhRvC..71b1301K}
C.~{Kurokawa} and K.~{Kat{\= o}}, ``{New broad {$0^{+}$} state in
  {$^{12}C$}},'' {\em Phys. Rev. C}, vol.~71, p.~021301, Feb. 2005.

\bibitem{2006PhRvC..74f4311A}
K.~{Arai}, ``{Resonance states of $^{12}C$ in a microscopic cluster model},''
  {\em Phys. Rev. C}, vol.~74, p.~064311, Dec. 2006.

\bibitem{2007PhRvC..76e4003T}
M.~{Theeten}, H.~{Matsumura}, M.~{Orabi}, D.~{Baye}, P.~{Descouvemont},
  Y.~{Fujiwara}, and Y.~{Suzuki}, ``{Three-body model of light nuclei with
  microscopic nonlocal interactions},'' {\em Phys. Rev. C}, vol.~76, p.~054003,
  Nov. 2007.

\bibitem{2007NuPhA.792...87K}
C.~{Kurokawa} and K.~{Kat{\= o}}, ``{Spectroscopy of {$^{12}C$} within the
  boundary condition for three-body resonant states},'' {\em Nucl. Phys. A},
  vol.~792, pp.~87--101, Aug. 2007.

\bibitem{1987PhRvC..36...54D}
P.~{Descouvemont} and D.~{Baye}, ``{Microscopic theory of the {$^{8}Be(\alpha,
  \gamma)^{12}C$} reaction in a three-cluster model},'' {\em Phys. Rev. C},
  vol.~36, pp.~54--59, July 1987.

\bibitem{2008JPhCS.111a2045O}
M.~{Orabi}, Y.~{Suzuki}, H.~{Matsumura}, Y.~{Fujiwara}, D.~{Baye},
  P.~{Descouvemont}, and M.~{Theeten}, ``{3{$\alpha$} description of $^{12}C$
  with microscopic nonlocal potentials},'' {\em J. Phys. Conf. Ser.}, vol.~111,
  p.~012045, May 2008.

\bibitem{2008PhLB..659..160S}
Y.~{Suzuki}, H.~{Matsumura}, M.~{Orabi}, Y.~{Fujiwara}, P.~{Descouvemont},
  M.~{Theeten}, and D.~{Baye}, ``{Local versus nonlocal {$\alpha$}{$\alpha$}
  interactions in a 3{$\alpha$} description of $^{12}C$},'' {\em Phys. Lett.
  B}, vol.~659, pp.~160--164, Jan. 2008.

\bibitem{2010JPhG...37f4010D}
P.~{Descouvemont}, ``{Three-{$\alpha$} continuum states},'' {\em J. Phys. G
  Nucl. Phys.}, vol.~37, p.~064010, June 2010.

\bibitem{2011PhRvC..83b4301Y}
T.~{Yoshida}, N.~{Itagaki}, and K.~{Kat{\= o}}, ``{Symplectic structure and
  monopole strength in {$^{12}C$}},'' {\em Phys. Rev. C}, vol.~83, p.~024301,
  Feb. 2011.

\bibitem{1998PhR...302..212M}
N.~{Moiseyev}, ``{Quantum theory of resonances: calculating energies, widths
  and cross-sections by complex scaling},'' {\em Phys. Rep.}, vol.~302,
  pp.~212--293, Sept. 1998.

\bibitem{1983PhR....99....1H}
Y.~K. {Ho}, ``{The method of complex coordinate rotation and its applications
  to atomic collision processes},'' {\em Phys. Rep.}, vol.~99, pp.~1--68, Oct.
  1983.

\bibitem{1995PhRvC..51..152A}
G.~S. {Anagnostatos}, ``{Alpha-chain states in {$^{12}C$}},'' {\em Phys. Rev.
  C}, vol.~51, pp.~152--159, Jan. 1995.

\bibitem{1992NuPhA.549..431M}
A.~C. {Merchant} and W.~D.~M. {Rae}, ``{Systematics of alpha-chain states in
  4N-nuclei},'' {\em Nucl. Phys. A}, vol.~549, pp.~431--438, Nov. 1992.

\bibitem{2007PThPh.117..655K}
Y.~{Kanada-En'yo}, ``{The Structure of Ground and Excited States of
  {$^{12}C$}},'' {\em Progr. Theor. Phys.}, vol.~117, pp.~655--680, Apr. 2007.

\bibitem{2004NuPhA.738..357N}
T.~{Neff} and H.~{Feldmeier}, ``{Cluster structures within Fermionic Molecular
  Dynamics},'' {\em Nucl. Phys. A}, vol.~738, pp.~357--361, June 2004.

\bibitem{1976JETP...43..205B}
A.~I. {Baz'}, ``{Diffusion-like processes in the quantum theory of
  scattering},'' {\em Soviet J. Exp. Theor. Phys.}, vol.~43, pp.~205--211, Feb.
  1976.

\bibitem{2010JPhG...37j5104M}
T.~{Mu{\~n}oz-Britton}, M.~{Freer}, N.~I. {Ashwood}, T.~A.~D. {Brown}, W.~N.
  {Catford}, N.~{Curtis}, S.~P. {Fox}, B.~R. {Fulton}, C.~W. {Harlin}, A.~M.
  {Laird}, P.~{Mumby-Croft}, A.~S.~J. {Murphy}, P.~{Papka}, D.~L. {Price},
  K.~{Vaughan}, D.~L. {Watson}, and D.~C. {Weisser}, ``{Search for the
  {$2^{+}$} excitation of the Hoyle state in {$^{12}C$} using the
  {$^{12}C(^{12}C,3\alpha)^{12}C$} reaction},'' {\em J. Phys. G Nucl. Phys.},
  vol.~37, p.~105104, Oct. 2010.

\bibitem{2009PhRvC..80d1303F}
M.~{Freer}, H.~{Fujita}, Z.~{Buthelezi}, J.~{Carter}, R.~W. {Fearick}, S.~V.
  {F{\"o}rtsch}, R.~{Neveling}, S.~M. {Perez}, P.~{Papka}, F.~D. {Smit}, J.~A.
  {Swartz}, and I.~{Usman}, ``{{$2^{+}$} excitation of the {$^{12}C$} Hoyle
  state},'' {\em Phys. Rev. C}, vol.~80, p.~041303, Oct. 2009.

\bibitem{2001PhRvC..63c4606V}
V.~{Vasilevsky}, A.~V. {Nesterov}, F.~{Arickx}, and J.~{Broeckhove},
  ``{Algebraic model for scattering in three-s-cluster systems. I. Theoretical
  background},'' {\em Phys. Rev. C}, vol.~63, p.~034606 (16 pp), Mar. 2001.

\bibitem{2001PhRvC..63c4607V}
V.~{Vasilevsky}, A.~V. {Nesterov}, F.~{Arickx}, and J.~{Broeckhove},
  ``{Algebraic model for scattering in three-s-cluster systems. II. Resonances
  in the three-cluster continuum of \makebox{$^{6}He$} and
  \makebox{$^{6}Be$}},'' {\em Phys. Rev. C}, vol.~63, p.~034607 (7 pp), Mar.
  2001.

\bibitem{2007JPhG...34.1955B}
J.~{Broeckhove}, F.~{Arickx}, P.~{Hellinckx}, V.~S. {Vasilevsky}, and A.~V.
  {Nesterov}, ``{The \makebox{$^5H$} resonance structure studied with a
  three-cluster \makebox{$J$}-matrix model},'' {\em J. Phys. G Nucl. Phys.},
  vol.~34, pp.~1955--1970, Sept. 2007.



\bibitem{A4Resonances2004}
V.~S. Vasilevsky, F.~Arickx, J.~Broeckhove, and V.~N. Romanov, ``Theoretical
  analysis of resonance states in \makebox{$^4H$}, \makebox{$^4He$} and
  \makebox{$^4Li$} above three-cluster threshold,'' {\em Ukr. J. Phys.},
  vol.~{\bf 49}, no.~11, pp.~1053--1059, 2004.

\bibitem{2001PhRvC..63f4604V}
V.~{Vasilevsky}, A.~V. {Nesterov}, F.~{Arickx}, and J.~{Broeckhove}, ``{S
  factor of the \makebox{$^{3}H(^{3}H,2n)^{4}He$} and
  \makebox{$^{3}He(^{3}He,2p)^{4}He$} reactions using a three-cluster exit
  channel},'' {\em Phys. Rev. C}, vol.~63, p.~064604 (8 pp), June 2001.

\bibitem{2004NuPhA.740..249K}
S.~{Korennov} and P.~{Descouvemont}, ``{A microscopic three-cluster model in
  the hyperspherical formalism},'' {\em Nucl. Phys. A}, vol.~740, pp.~249--267,
  Aug. 2004.

\bibitem{2009PhRvC..80d4310D}
A.~{Damman} and P.~{Descouvemont}, ``{Three-body continuum states in a
  microscopic cluster model},'' {\em Phys. Rev. C}, vol.~80, p.~044310, Oct.
  2009.

\bibitem{2002PThPh.107..993F}
Y.~{Fujiwara}, Y.~{Suzuki}, K.~{Miyagawa}, and {Michio}, ``{Redundant
  Components in the 3{$\alpha$} Faddeev Equation Using the 2{$\alpha$} RGM
  Kernel},'' {\em Prog. Theor. Phys.}, vol.~107, pp.~993--1000, May 2002.

\bibitem{2009NuPhA.826...24L}
Y.~A. {Lashko} and G.~F. {Filippov}, ``{The role of the Pauli principle in
  three-cluster systems composed of identical clusters},'' {\em Nucl. Phys. A},
  vol.~826, pp.~24--48, July 2009.

\bibitem{1965AnPhy..35...18Z}
W.~{Zickendraht}, ``{Construction of a complete orthogonal system for the
  quantum-mechanical three-body problem},'' {\em Ann. Phys.}, vol.~35,
  pp.~18--41, Oct. 1965.

\bibitem{Nyiri:1979qm}
J.~Nyiri and Y.~A. Smorodinsky, ``{Symmetrical basis in the three-body problem.
  (In russian)},'' {\em Yad. Fiz.}, vol.~29, pp.~833--844, 1979.

\bibitem{2010PPN....41..716N}
A.~V. {Nesterov}, F.~{Arickx}, J.~{Broeckhove}, and V.~S. {Vasilevsky},
  ``{Three-cluster description of properties of light neutron- and proton-rich
  nuclei in the framework of the algebraic version of the resonating group
  method},'' {\em Phys. Part. Nucl.}, vol.~41, pp.~716--765, Sept. 2010.

\bibitem{kn:Minn_pot1}
D.~R. Thompson, M.~LeMere, and Y.~C. Tang, ``Systematic investigation of
  scattering problems with the resonating-group method,'' {\em Nucl. Phys.},
  vol.~{\bf A286}, no.~1, pp.~53--66, 1977.

\bibitem{2005PhRvC..71d4322S}
A.~{Sytcheva}, F.~{Arickx}, J.~{Broeckhove}, and V.~S. {Vasilevsky},
  ``{Monopole and quadrupole polarization effects on the {$\alpha$}-particle
  description of \makebox{$^8Be$}},'' {\em Phys. Rev. C}, vol.~71, p.~044322,
  Apr. 2005.

\bibitem{kn:wilderm_eng}
K.~Wildermuth and Y.~Tang, {\em A unified theory of the nucleus}.
\newblock Braunschweig: Vieweg Verlag, 1977.

\end{thebibliography}

\end{document}